\documentclass[journal]{IEEEtran}
\usepackage[T1]{fontenc}
\usepackage{amsmath,bm,amsfonts,amssymb,graphicx,color,multirow,setspace,xfrac,comment}
\usepackage{booktabs}
\usepackage{epstopdf}
\usepackage{mathtools}
\usepackage{lipsum}
\usepackage{cite,citesort}
\usepackage{float}
\usepackage{balance}
\usepackage{gensymb}
\usepackage[acronym]{glossaries}
\begin{document}
\title{Standardization of Propagation Models: 800 MHz to 100 GHz -- A Historical Perspective\vspace{2pt}}
\author{Harsh~Tataria,~\IEEEmembership{Member,~IEEE}, 
Katsuyuki~Haneda,~\IEEEmembership{Member,~IEEE}, 
Andreas~F.~Molisch,~\IEEEmembership{Fellow,~IEEE}, 
Mansoor~Shafi,~\IEEEmembership{Life~Fellow,~IEEE}, and 
Fredrik~Tufvesson,~\IEEEmembership{Fellow,~IEEE}
\thanks{H.~Tataria and F.~Tufvesson are with the Department of Electrical and Information Technology, Lund University, Sweden (e--mail:\{harsh.tataria, fredrik.tufvesson\}@eit.lth.se).}
\thanks{K.~Haneda is with the School of Engineering, Aalto University, Espoo, Finland (e--mail: katsuyuki.haneda@aalto.fi).}
\thanks{A. F. Molisch is with the Department of Electrical Engineering, University of Southern California, USA (e--mail: molisch@usc.edu). His work was supported by the National Science Foundation (NSF) and the National Institute of Standards and Technology (NIST).}
\thanks{M. Shafi is with Spark New Zealand, Wellington, New Zealand (e--mail: mansoor.shafi@spark.co.nz).}\vspace{-2pt}}
\IEEEspecialpapernotice{Invited Paper\vspace{-13pt}}
\maketitle

\begin{abstract}
Propagation models constitute a fundamental building block of wireless communications research. Before we build and operate real systems, we must understand the science of radio propagation, and develop channel models that both reflect the important propagation processes and allow a fair comparison of different systems. In the past five decades, wireless systems have gone through five generations, from supporting voice applications to enhanced mobile broadband. To meet the ever increasing data rate demands of wireless systems, frequency bands covering a wide range from 800 MHz to 100 GHz have been allocated for use. The standardization of these systems started in the early/mid 1980's in Europe by the European Telecommunications Standards Institute with the advent of Global System for Mobile Communications. This motivated the development of the first standardized propagation model by the European Cooperation in Science and Technology (COST) 207 working group. These standardization activities were continued and expanded for the third, fourth, and fifth generations of COST, as well as by the Third Generation Partnership Project, and the International Telecommunication Union. This paper presents a historical overview of the standardized propagation models covering first to fifth--generation systems. In particular, we discuss the evolution and standardization of pathloss models, as well as large and small--scale fading parameters for single antenna and multiple antenna systems. Furthermore, we present insights into the progress of deterministic modelling across the five generations of systems, as well as discuss more advanced modelling components needed for the detailed simulations of millimeter--wave channels. A comprehensive bibliography at the end of the paper will aid the interested reader to dig deeper.
\end{abstract}

\vspace{-3pt}

\vspace{-11pt}
\section{Introduction}
\label{Introduction}
Humanity has been interested in communication since the world began. The discovery of wireless (a.k.a. radio) communications has helped people to communicate over large physical distances using the wireless medium. The author of \cite{roots} states that the persons who can legitimately be called the \emph{``fathers of wireless communications"}, can be divided into two groups: (1) The discoverers and (2) The inventors. The discoverers are Michael Faraday (1791-1867) William Thomson (Lord Kelvin) (1824-1907), James Clerk Maxwell (1831-1879), Heinrich Rudolf Hertz (1857-1894), and Jagdish Chandra Bose (1858-1937); the inventors are Nikola Tesla (1856-1943), Guglielmo Marconi (1874-1937), Alexander Popov (1859-1906), Reginal Aubrey Fessenden (1866-1932), and Lee De Forest (1873-1961). Those and many others designed electrical circuits that could transmit and receive information by means of electromagnetic waves, changing the way we live. A key in this revolution has been to develop an understanding of the science of radio propagation \cite{GREENSTEIN1}, which fundamentally tells us \emph{how} radio waves propagate through different environments, and what type of system performance we can consequently expect. Unless we understand radio propagation, we simply cannot design, build and deploy radio systems, let alone harvest their great benefits. While Maxwell's equations are universal, they are too complex for practical use; yet useful approximations of propagation channels depend on the frequency range and the system under consideration, which have dramatically changed over the past 40 years. So, in the words of the late Larry Greenstein (1937-2018), a prominent figure in the field of radio propagation (referred to by many as the father of modern radio propagation research) \cite{GREENSTEIN2}, \emph{``Every time a new system has been built in a new band, in a new environment, or for a new service, major questions have had to be answered about the nature of the radio propagation. It was true for Marconi's wireless telegraph; it is true for today's cellular systems; and it will be true for as long as people dream up new ways to use radio waves. Propagation is different at 6 GHz than at 850 MHz; indoor propagation differs from outdoor propagation; fixed wireless paths differ from mobile ones; and so on"} \cite{GREENSTEIN1}. We will add to this and say that wave propagation at 100 GHz is very different from propagation below 6 GHz. 

Propagation studies are time consuming and expensive. Understanding nature by means of models, no matter how complex they are, is not a trivial task. Channel models are based on deterministic and stochastic parameters -- all derived from extensive measurements over a long period of time. Standardized channel models enable a smoother transition in understanding the laws of nature and are -- if done well -- the result of extensive field measurements conducted worldwide by academia, industry and other research scientific personnel. Standardization of propagation channel models are essential for the development of commercial radio hardware that complies with \emph{common} specifications. They enable the evaluation of different candidate systems under a common framework of channel models. This is essential to take advantage of global harmonization and economies--of--scale. These models thus play a crucial role in the development of many physical layer functions such as modulation and coding techniques, multiple access techniques, signal processing methods, transceiver architectures, antennas and antenna arrays, methods for spectral efficiency and performance improvement, etc.\footnote{Instruments known as \emph{fading simulators} are also now available to play back standardized channel impulse responses so that commercial hardware and its sub--systems can be pre--tested in a laboratory environment before being deployed out in the field.}  To this end, it would be misguiding to assume that propagation models are just a collection of random numbers, and any distribution of a parameter will suffice; rather a strong connection to the propagation characteristics in the envisioned deployment areas, and to the characteristics of the systems is required. In many research papers, the generation of the wireless channel is based on some kind of wave propagation model published either by the standardization bodies or derived from measurements. \emph{Unlike others, the aim of this paper is to present a historical overview on the evolution of standardized propagation models over five generations of wireless systems.} 

It is well known that radio systems have undergone a generational change every ten years. The first generation (1G) of systems in the United States (US), known as Analog Mobile Phone Systems (AMPS), started in 1974 when the Federal Communications Commission (FCC), the frequency regulator in the US, allocated 40 MHz of bandwidth in the 800--900 MHz frequency region. This is now known as Third Generation Partnership Project (3GPP) band 5 \cite{36104}. 1G systems were aimed to provide voice calling services only and there were different systems in different countries/regions, e.g., the Nordic Mobile Telephone (NMT) system in the Nordic countries, Total Access Communications System (TACS) in the United Kingdom and Radio Telephone Network C (C--450) in West Germany. The bandwidth allocated to each user in these systems was typically 20--30 kHz. From a wave propagation viewpoint, \emph{received power} characterization due to pathloss was the only property of interest, see e.g., \cite{cox1,cox2,cox3,suzuki} (though some of these references also measured power--delay profiles). After this, in the 1980s, second--generation (2G) mobile communication systems were introduced in 3GPP band 5, as well as other bands in Europe. The Global System for Mobile Communications (GSM) was standardized in the early 1980's spurred by major research efforts within European Union (EU) consortia \cite{steele,Molisch1}. With a carrier bandwidth of 200 kHz, GSM became a dominant standard in the world for radio services. Besides voice, low--rate data capability up to tens (later hundreds) of kilobits--per--second (kbps) was also provided. Consequently, it became important to study aspects of \emph{delay dispersion} of the radio channel. Seminal measurements were carried out by Cox, Rustako, Greenstein, and others \cite{cox1,cox2,cox3,RUSTAKO1,GOLDSMITH1} under various different environments in New York city and New Jersey, US.  

The standardization of wave propagation models did not happen until 2G mobile systems, where the GSM specifications relied on the European Cooperation in Science and Technology (COST) 207 model \cite{COST207}. The International Telecommunication Union (ITU) standardization of channel models did not happen until third--generation (3G) of mobile systems, a.k.a. International Mobile Telecommunications (IMT)--2000, which were introduced in 1997 \cite{ITUR,Molisch1}. The standardization of IMT--2000 required a method to evaluate candidate 3G technologies, and for this purpose, a propagation model had to be standardized \cite{ITUR}. Parallel developments towards standardization of 3G systems were taking place by the 3GPP. These systems initially operated in 3GPP band 1--i.e., the 2 GHz band, but later were also deployed in many other bands (for exact band numbers and ranges the reader is referred to \cite{36104}). 3G systems had a bandwidth of 5 MHz and offered peak data rates of 2 Mbps focusing on voice and early multimedia applications, thus improved delay resolution was required. Advanced versions of 3GPP also started 
to consider the \emph{spatial} domain to increase peak data rates, thanks to the seminal contributions of Winters, Foschini, Gans, Teletar and others \cite{WINTERS1,FOSCHINI1,TELETAR1,TAROKH1}. This motivated the development of the COST 259 model \cite{COST259book,Molisch259,Asplund259} in 2000 and the 3GPP Spatial Channel Model (SCM) in 2003 \cite{3GPP25996,CALCEV1}. Yet another five years later, fourth--generation (4G) systems, a.k.a. IMT--Advanced or 3GPP Release 8 emerged in 2010. 
Early 4G systems deployed two transmit and two receive antennas, known as $2\times{}2$ multiple--input multiple--output (MIMO). They were capable of offering peak rates of 150 Mbps within a bandwidth of 20 MHz. Building on the SCM model structure and extensive measurements in the The Wireless World Initiative New Radio (WINNER) project of the EU, propagation models for 4G systems were standardized by the 3GPP in \cite{3GPP25996}, and the ITU Radio Communication Sector (ITU--R) in \cite{ITURM2135}. The latter provided guidelines for the evaluation of IMT--Advanced systems. Driven by the further increase in data rate demands, later releases of 4G systems featured enhancements in the MIMO order. Using 4 port cross--polarized antennas, $4\times{}4$ MIMO systems were deployed \cite{Kozono}. Together with this, the introduction of two--dimensional antenna arrays enabled the deployment of multiuser MIMO systems \cite{NAM1}. This facilitated the necessity of obtaining the \emph{full--dimensional} (a.k.a. three--dimensional) nature of the channel, to accurately model the multipath amplitudes, delays, azimuth/zenith angles--of--departure (AODs/ZODs)\footnote{In line with the standardization terminology, we refer to the elevation domain of the radio channel as the \emph{zenith} domain.}, azimuth/elevation angles--of--arrival (AOAs/ZOAs), polarization and Doppler parameters \cite{3GPP3}. 

Recently, fifth--generation (5G) systems, known as IMT--2020 or 3GPP Release 15 New Radio (NR) are in early stages of deployment. Live commercial networks are in place in various parts of North America and Asia, gradually extending towards Europe and Oceanic regions \cite{SHAFIICC2020}. IMT--2020/3GPP NR systems are the first to operate across a wide range of \emph{multiple} frequency bands. Depending on the country, they are designed to operate within the \emph{C--band}, i.e., 3GPP band numbers N77 (3300--4200 MHz), N78 (3300--3800 MHz) and N79 (4400--5000 MHz), as well as in bands approaching \emph{millimeter-wave} (mmWave) frequencies, i.e., 3GPP band N257 (26.5--29.5 GHz) and N258 (24.25-27.5 GHz), respectively. The N258 band, and other high frequency bands up to 100 GHz were officially granted at the World Radio Communications Conference 2019 (WRC--19) \cite{WRC2019}, and later releases of 5G--NR, such as 3GPP Release 16 have recently disclosed plans to extend the frequency band of operation to beyond 52 GHz \cite{3GPPTR21916}. 5G systems are designed to provide peak data rates of up to 20 Gbps--see e.g., \cite{IMTVision} and \cite{SHAFI1} for a complete list of NR performance parameters. 3GPP Release 15 facilitates bandwidths of up to 100 MHz for bands below 6 GHz and up to 400 MHz for bands approaching mmWaves.\footnote{In this paper, we employ common notation from the literature and refer to 3GPP bands N257 and N258 as mmWave bands, even though they do not fall truly in the mmWave regime,  which ranges from 30 GHz to 300 GHz.} For frequencies below 6 GHz, in order to maintain uniformly good service while retaining wide area coverage, the use of large antenna arrays, a.k.a., massive MIMO has been proposed. Here, aggressive spatial multiplexing due to simultaneous service of many mobile stations (MSs) within the same time--frequency resource is possible \cite{MARZETTA1,LARSSON1}. For bands above 6 GHz, massive MIMO is now an essential technique used to provide the necessary \emph{array gain}, closing the link budget to communicate over moderate distances \cite{SHAFI2,RAPPAPORT1}. The development of propagation channel models for 5G--NR/IMT--2020 systems has undergone several phases in 3GPP and ITU--R, resulting in the standards \cite{3GPPTR38900,3GPP2,ITU2}. The standards have also identified propagation features influenced by operation at higher frequency bands, over wider bandwidths, such as blockage modelling, outdoor--to--indoor penetration loss, and oxygen/molecular absorption losses. In addition to this, spatial consistency modelling is defined in \cite{3GPP2,ITU2} for the first time, although a long history of spatial consistency modelling exists for the COST models.

In light of the above discussions, it is clear that bandwidths have evolved from 30 kHz to 400 MHz and peak data rates have evolved from 200 kbps to 20 Gbps, complemented by the phenomenal advances in MIMO order. As we increased the carrier frequency, carrier bandwidths and data rate requirements, the complexity of standardized wave propagation models has also significantly increased. In this paper, we present the evolution of  propagation models in the respective standards from 1G to 5G. We assume that readers have basic familiarity with the terminology used in propagation models and measurements literature. Alternatively, readers are first referred to the related discussions in, e.g., the textbooks
\cite{Molisch1,GOLDSMITHB,RAPPAPORT2}. Irrespective of the generation, frequency of operation and bandwidth, all propagation models try and capture small and large--scale fading variations. These are briefly discussed below for the interest of setting up the discussion in the following sections, and for completeness purposes: 
\begin{itemize}
    \item  \emph{Large--scale fading:} These are the fading effects captured over a large spatial area (typically hundreds of wavelengths), and henceforth denote the variation of the \emph{mean} received signal power. Most often caused by \emph{shadowing} due to large objects blocking the transmitted waveform en--route to the receiver, large--scale fading is often described by a lognormal distribution (equivalently by a zero--mean Gaussian distribution when the variables are on a decibel (dB) scale). The standard deviation of this power fluctuation process is known as the shadow fading standard deviation. Given a propagation environment, the large--scale mean power itself experiences distance--based attenuation, and is usually captured in pathloss models.  
    \item \emph{Small--scale fading:} small--scale fading occurs on a much smaller spatial scale (on the order of a wavelength or even less). Here the received signal power undergoes rapid fluctuations due to the \emph{superposition} of multipath components (MPCs), constructively adding or destructively cancelling each other. Therefore, all the parameters that characterize the MPCs, such as delays, AOD/AOAs, ZOD/ZOAs are deemed as small--scale fading parameters. For a large number of incoming wavefronts, the overall amplitude of the superposed MPCs is usually modelled as a complex Gaussian random process, whose magnitude obeys a Rayleigh distribution; hence the name \emph{Rayleigh fading}. However, when a dominant component, such as line--of--sight (LOS), is also present in addition to many smaller MPCs, a Ricean distribution is used to better describe the amplitude distribution \cite{greensteinmoment,TATARIA1,TATARIA2} that is characterised by a Rice factor, $K$. When $K$ is small enough (so that the LOS power is similar to other components), the Ricean distribution converges to a Rayleigh distribution \cite{TATARIA1,TATARIA2}, whilst for large values of $K$, the Ricean distribution converges towards a Gaussian distribution centered around the amplitude of the dominant MPC. In the standardized channel models, $K$ is a random variable whose parameters are dependent on the carrier frequency. 
      \end{itemize}
The above discussion captures the essence of simple statistical fading models. In addition to the above, there are other parameters of a channel model that do not fit in the above categories: examples are oxygen absorption, probability of LOS, outdoor--to--indoor penetration losses, human blocking, and spatial consistency. Additionally, there are also antenna and system--related parameters like cross--polarization discrimination factors that are needed when modelling wave propagation with cross--polarized antenna elements \cite{Kozono}, as for 4G and 5G systems. The above parameters are needed to accurately characterize propagation for a given environment, and hence are of interest to standardization bodies. 

The organization of this paper is as follows: We first describe pathloss models from 1G to 5G systems. Here we show how pathloss increases when operating frequency increases from sub 1 GHz bands all the way to 100 GHz. In order to  mitigate the high pathloss, antenna arrays are utilized. However, in doing so, the beamwidth decreases and antennas become highly directional. As a consequence, we discuss \emph{directional} pathloss models, particularly in the context of 5G--NR systems. This is followed by a section on impulse response evolution, including \emph{non--directional} impulse response models to \emph{double--directional} models; the latter also includes the spatial channel impulse response needed for the simulation of conventional MIMO and massive MIMO channels. A special emphasis is placed on describing the impulse response evolution over the various generations of wireless systems. Following this, \emph{map--based} and \emph{ray tracer--based} deterministic models are discussed. These quasi--deterministic models use geometric information of a cellular site as a countermeasure to statistical models. The discussion of additional modelling components is then presented, followed by a comprehensive bibliography for the interested reader to delve deeper into this field. 

\section{Pathloss Models}
\label{pathlossmodels}
\vspace{-1pt}
Radio signals launched by a transmitter experience signal attenuation as they traverse through the propagation channel. This attenuation is a function of the carrier frequency, heights of the base station (BS) and MS, BS--MS link distance, as well as the environment--type (i.e., dense urban, suburban, rural etc). Pathloss is a large--scale fading parameter, which determines the \emph{mean} signal attenuation as a function of the BS to MS distance \cite{Molisch1}. At a given link distance, there are slow variations of up to 10 dB around the mean value, over a spatial scale of hundreds of wavelengths. These variations are due to man made and natural objects such as tunnels, hills, buildings, etc, and are captured by shadow fading. The exact structure of shadowing is dependent on the geometry surrounding the BS and MS, as well as the operating frequency. To this end, the shadow fading statistics are important parameters for accurate estimation of the received power at a given MS location. Assuming for the sake of the argument that both the BS and MS are separated by a distance $d$ meters (m) in free--space, a transmitted signal with frequency $f$ (corresponding to wavelength $\lambda$) having transmit power $P_{\textrm{t}}$ yields a received power which is well characterized by the  Friis' equation \cite{Molisch1}:  
\vspace{-4pt}
\begin{equation}
\label{Friisformula}
P_{\textrm{r}}=P_{\textrm{t}}\hspace{2pt}
G_{\textrm{t}}\hspace{2pt}G_{\textrm{r}}
\left(\frac{\lambda}{4\pi{}d}\right)^{\hspace{-1pt}2}.
\vspace{-2pt}
\end{equation}
Here, $G_{\textrm{t}}$ and $G_{\textrm{r}}$ denote the transmit and 
receive antenna gains, and $\left(\lambda/4\pi{}d\right)^{2}$ is the so--called \emph{free-space loss factor}. Several important assumptions exist in the formulation of the free--space equation: Firstly, holding all other variables constant, if $\lambda$ decreases, \emph{implying that $f$ increases}, power received \emph{decreases quadratically}. Nevertheless, this comes with a caution that the antenna gains ($G_{\textrm{t}}, G_{\textrm{r}}$) at the both link ends are kept \emph{constant} (i.e., fixed) with increasing $f$, like a half--wave dipole. It is well known that there is a straightforward relationship between the antenna gain and its \emph{effective area}. For instance, taking the example of the transmit antenna with gain $G_{\textrm{t}}$, its effective area can be computed by  $A_{\textrm{t}}=(\lambda^2/4\pi)\hspace{1pt}\hspace{1pt}
G_{\textrm{t}}$, or equivalently
$G_{\textrm{t}}=A_{\textrm{t}}\hspace{1pt}(4\pi/\lambda^2)$. From such formulation, it can be readily seen that if $\lambda$ decreases, with a \emph{constant} effective area, the antenna gain \emph{increases} with the square of $\lambda$. \emph{To this end, if one is willing to invest the same effective area in the antenna irrespective of $f$, the antenna gains will increase leading to a decrease in the pathloss, and hence an increase in the received power.} Note that increasing the number of antenna elements for a constant \emph{physical area} in an array will result in heavier antennas (due to more electronic components) that may pose tower and wind loading problems, as well as power consumption issues. This is especially true for 5G systems, where a large number of BS elements are interfaced with dedicated radio frequency up/down--conversion chains for implicit and explicit beamforming \cite{SHAFIICC2020}. Naturally, the validity of \eqref{Friisformula} is restricted to the \emph{far--field} of the antenna -- i.e., the transmit and receive antennas have to be at least one (to ten or larger) Fraunhofer distance(s) away. The readers are asked to refer to \cite{Molisch1,RAPPAPORT2} for further discussions.  
\vspace{-7pt}
\subsection{Pre-2G and 2G Pathloss Models}
\label{Pre2GPand2GPathlossModels}
\vspace{-1pt}
Systems prior to 2G and early 2G systems were designed using empirical pathloss models. The seminal work of Okumura and Hata in \cite{OKUMURAHATAMODEL} presents one such a model\footnote{An extension of the Okumura--Hata model to the 2 GHz band is sometimes also referred to also as the COST Hata model (see e.g., \cite{COST231}).}. The general structure of the model expresses the pathloss on the dB scale for a 2D link distance $d$ as 
\begin{equation}
    \label{OkumuraHataModelStructure}
    \textrm{PL}^{\textrm{Pre--2G/2G}}_{\textrm{Okumura--Hata}}=
    A+B\log_{10}\left(d\right)+C, 
    \vspace{-2pt}
\end{equation}
where $A$, $B$ and $C$ are parameters which depend on the carrier frequency, environment, and relative antenna heights\footnote{Note that $\log_{10}\left(\cdot\right)$ is referred to as 
$\log\left(\cdot\right)$ unless otherwise specified in the text. Also, we note that $d$ denotes the 2D  link distance.}. Generally speaking, the parameter $A$ increases with carrier frequency and decreases with increasing height of the BS and MS. The pathloss exponent, $B/10$,  decreases with increasing height of the BS. The model is valid for frequencies up to 1500 MHz only and is intended for studies involving large cells with the BS being placed higher than the surrounding rooftops. The study in \cite{COST207} presents the model for urban areas where 
\vspace{-5pt}
\begin{align}
    \nonumber
    \textrm{PL}^{\textrm{Pre--2G/2G}}_{\textrm{COST 231}}=\hspace{3pt}&69.55+26.16\log(f)-13.82\log(h_{\textrm{t}})-a(h_{\textrm{r}})\\
    \label{OkumuraHataPathloss}&+44.9-6.55\log(h_{\textrm{t}})\log(d),
\end{align}
where $f$ is the carrier frequency in Hz, and $h_{\textrm{t}}$, $h_{\textrm{r}}$ are transmitter and receiver heights in meters. For a moderately sized city, 
\vspace{-5pt}
\begin{equation} 
    \label{OkumuraHataPathloss2}
    a\left(h_{r}\right)=\left(1.1\log(f)-0.7\right)h_{\textrm{r}}-1.56\log\left(f\right)-0.8. 
    \vspace{-2pt}
\end{equation}
Subsequent to this, other models were developed such as the COST 231 Walfish--Ikegami model \cite{WALFISCH1,OKOMOTO1} and the Motely--Keenan model \cite{MOTLEYKEENAN1}, respectively. The former model is also suitable for microcells and small macrocells, since it has fewer restrictions on the distance between the BS and MS, as well as antenna heights. The latter model includes the effects of floor and wall penetration as constants which depend on the operating frequency. The COST 231 Walfish--Ikegami model characterizes the total pathloss as a function of the free--space loss, multi--screen loss along a propagation path due to a series of rooftops perpendicular to the propagation path, as well as attenuation from the last roof--edge to the MS. The model assumes a Manhattan street grid (i.e., streets intersecting at right angles), constant building heights and uniform terrain. Note that the model only considers over--rooftop propagation and does not include the effects of waveguiding through street canyons, which may lead to an underestimation of the received power. Depending on the environment, pre--2G and 2G systems most often characterized pathloss exponents between 3 to 4. 
\vspace{-10pt}
\subsection{Pathloss Models for 3G Systems}
\label{PathlossModelsfor3GSystems}
Pathloss models for 3G systems were standardized in \cite{ITUR,BERTONI1}\footnote{Reference \cite{ITUR} was a result of many input contributions made to ITU--R Task Group (TG) 8/1 by the ITU--R members.} using the well known work of Bertoni and co-workers as a basis. Pathloss models were divided into three environments of \emph{indoor offices}, \emph{outdoor--to--indoor} and \emph{vehicular} test  environments. The pathloss model for the \emph{indoor office} environment follows a simplified form, which is derived from the COST 231 indoor model. The model is expressed as
\vspace{-3pt}
\begin{equation}
    \label{IndoorPLModel3G}
    \textrm{PL}^{\textrm{3G}}_{\textrm{Indoor}}=37+30\log\left(d\right)+18.3\hspace{1pt}n^{\big[
    \left(n+2/n+1\right)-0.46\big]}, 
    \vspace{-1pt}
\end{equation}
where $d$ is the link separation distance in km and $n$ denotes the number of building floors in the path. Unlike other indoor models, here the lognormal shadow fading standard deviation is inclusive of the variations in floor penetration losses, and hence a value of 12 dB is quoted in \cite{ITUR}. Alternatively, the pathloss model for outdoor--to--indoor environments is given by 
\vspace{-7pt}
\begin{equation}
    \label{OutdoortoIndoorModel3G}
    \textrm{PL}^{\textrm{3G}}_{\textrm{Outdoor-to-Indoor}}=40\log\left(d\right)+30\log\left(f\right)+49,
\end{equation}
where $d$ is the 2D link distance in kilometer (km) and $f$ is the carrier frequency of 2000 MHz for IMT--2000 band applications. The model describes the worst case propagation behavior. On top of \eqref{OutdoortoIndoorModel3G}, lognormal shadow fading with a standard deviation of 10 dB for outdoor 
users and 12 dB for indoor users is assumed. The average building penetration loss is assumed to be 12 dB with a standard deviation of 8 dB. The pathloss model for the vehicular test environment in urban and suburban areas outside the high rise core where the buildings are of nearly uniform height follows 
\begin{align}
    \nonumber
    \textrm{PL}^{\textrm{3G}}_{\textrm{Vehicular}}=&40\left(1-4\times{}10^{-3}\Delta{}h_{\textrm{t}}\right)\log\left(
    d\right)-18\log\left(\Delta{}h_{\textrm{t}}\right)\\
    \label{VehicularModel3G}
    &+21\log\left(f\right)+80. 
\end{align}
Note that $\Delta{}h_{\textrm{t}}$ denotes the BS height in m, measured from the average rooftop level and $d,f$ are as defined previously for the indoor--to--outdoor model. Although the model is valid for a range of $\Delta{}h_{\textrm{t}}=\{0,\dots,50\}$ m, the BS height is typically assumed to be fixed at 15 m above the average rooftop, such that $\Delta{}h_{\textrm{t}}=15$ m. The pathloss model is accompanied by a lognormal shadow fading model with 10 dB standard deviation for both urban and suburban areas. An important observation from the three aforementioned environments is the fact that the outdoor--to--indoor and vehicular test cases have a direct frequency dependent pathloss component in $\log\left(f\right)$, unlike the indoor model in \eqref{IndoorPLModel3G}. Naturally, all models exhibit a log--distance relationship as seen by the relevant expressions as well as dependency of pathloss on the antenna height.
\vspace{-9pt}
\subsection{Pathloss Models for 4G Systems}
\label{PathlossModelsfor4GSystems}
With the emergence of MS--specific full--dimensional beamforming, the standardization of 4G systems (IMT--Advanced) led to the development of more in--depth 3D pathloss models in \cite{ITURM2135} and \cite{3GPP3}. The models standardized by the ITU--R and 3GPP are applicable from 450 MHz to 6 GHz, and are tailored to several physical settings based on knowledge of the available measurement literature, see e.g, \cite{WINNER}. More specifically, \emph{four} usage scenarios are discussed in \emph{3D urban macrocellular (UMa), microcellular (UMi)}, and \emph{UMa/UMi outdoor--to--indoor}, respectively. The UMi scenarios assume that the BS height is below the surrounding rooftop heights, while the UMa scenario assumes that the BS is above the surrounding buildings. For each of these scenarios, two sub--cases for LOS and NLOS pathloss models are provided in \cite{3GPP3}. The 3D UMi LOS pathloss model is given by: 
\vspace{-1pt}
\begin{equation}
\label{4GPathlossModel1}
\textrm{PL}_{\textrm{3D UMi LOS}}^{\textrm{4G(a)}}= 
22.0\log\left(d\right)+28.0+20\log\left(f\right), 
\vspace{-1pt}
\end{equation}
where $d$ is the 3D link distance from the BS to MS and $f$ is the carrier frequency in GHz. Note that the model assumes a height of the BS to be 10 m or smaller, and the height of the MS is in between 1.5 m and 22.5 m. The applicable distance range of the model is when $10$ m $\leq{}d\leq{}d_{\textrm{break}}$, where $d_{\textrm{break}}$ is the break--point distance. Beyond the break--point distance, up to 5000 m, the pathloss is given by 
\vspace{-3pt}
\begin{align}
    \nonumber  
    \textrm{PL}^{\textrm{4G(b)}}_{\textrm{3D UMi LOS}}& =  40.0\log\left(d\right)+28.0+20\log\left(f\right)\\ \label{4GPathlossModel2}
    &-9\log\hspace{1pt}[\left(d_{\textrm{break}}\right)^{2}+
    \left(h_{\textrm{t}}-h_{\textrm{r}}\right)^{2}]. \\[-18pt]
    &\nonumber
\end{align}
With a fixed BS height of 10 m, the equivalent NLOS pathloss under the same scenario is given by 
\vspace{-3pt}
\begin{align}
   \nonumber
    \textrm{PL}^{\textrm{4G}}_{\textrm{3D UMi NLOS}}& = 
    36.7\log\left(d\right)+22.7+26\log\left(f\right)\\  \nonumber
   &-0.3\left(h_{\textrm{r}}-1.5\right). \\[-21pt] 
   & \label{4GPathlossModel3}
\end{align}
The pathloss model has a maximum modelling 3D distance range of 2000 m. \emph{Rather interestingly, the 3GPP modelling methodology describes the pathloss model for 3D UMa environment in LOS situations to be the same as that for the 3D UMi LOS case in \eqref{4GPathlossModel1} and \eqref{4GPathlossModel2}}. Moreover, the pathloss model for 3D UMa NLOS case is given by 
\vspace{-2pt}
\begin{align}
    \nonumber
    \textrm{PL}_{\textrm{3D UMa NLOS}}^{\textrm{4G}}& =  161.04-7.1\log\left(W\right)+7.5\log\left(h\right)\\ \nonumber
    &-\left[24.37-3.7\left(h/h_{\textrm{t}}\right)^{2}\right]\log\left(h_{\textrm{t}}\right)\\ \nonumber
    &+\left[\hspace{1pt}43.42-3.1\log\left(h_{\textrm{t}}\right)\hspace{1pt}\right]\left[\hspace{1pt}\log\left(d\right)-3\hspace{1pt}\right]\\\nonumber
    &+20\log\left(f\right)-\left\{3.2\left[\hspace{1pt}
    \log\left(17.625\right)\hspace{1pt}\right]^{2}-4.97\right\}\\ \label{4GPathlossModel4}
    &-0.6\left(h_{\textrm{r}}-1.5\right). \\[-17pt]
    &\nonumber
\end{align}
Note that $W$ \emph {denotes} the street width, $h$ \emph {denotes} the average building height. The applicability range of the parameters involved in \eqref{4GPathlossModel4} are as follows: $5$~m$~<h<50$~m, $5$~m~$<W<50$~m, $10$~m $<h_{\textrm{t}}<150$~m, and 
$1.5$~m$ <h_{\textrm{r}}<22.5$~m. The 3D UMi and UMa \emph {outdoor--to--indoor} pathloss models follow the same structure as their outdoor counterparts with the addition of extra wall loss (related to the electrical thickness of the wall) and loss inside the building \cite{3GPP3} scenario. Due to this reason, we omit listing out the equations for this case, as interested readers can refer to \cite{3GPP3}. For all of the aforementioned 4G pathloss models, shadowing is modelled as a zero--mean lognormal process in the dB domain (as for 3G pathloss models), where its standard deviation is varying from 3 dB in UMi LOS, 4 dB in UMi NLOS and UMa LOS, 6 dB in UMa NLOS, up to 7 dB in UMi/UMa outdoor--to--indoor environments \cite{3GPP3,ITURM2135}. 

\vspace{-10pt}
\subsection{Pathloss Models for 5G Systems}
\label{PathlossModelsfor5GSystems}
\begin{figure}[!t]    
\vspace{-12pt}
    \hspace{-2pt}
    \includegraphics[width=9cm]{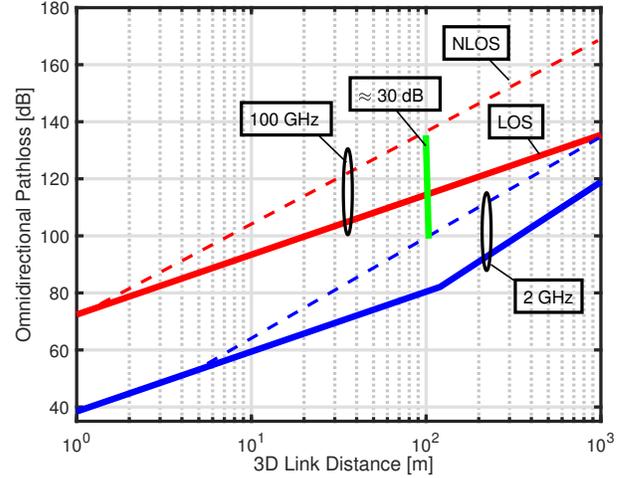}
    \caption{3GPP TR 38.901 standardized pathloss model for the 3D UMa LOS and NLOS scenarios across 2 GHz and 100 GHz. The NLOS cases at both frequencies are depicted in dashed line, while the solid lines are used to denote LOS cases. For further information, see \cite{3GPP2}.}
    \label{PathlossComparison}
\end{figure}

The pathloss models for IMT--2020 (5G--NR) systems are described in \cite{3GPP2} and \cite{ITU2}. The model in \cite{ITU2} consists of input contributions from all the members of ITU--R Working Party (WP) 5D. For example, the indoor hotspot model is based on measurements in \cite{Pan}, measurements for other environments came from multiple organisations to ITU--R WP 5D. These models are also a result of a collective efforts of many propagation measurements by industrial and academic organizations reported in \cite{GLOBECOMWP1,GLOBECOMWP2}. As mentioned in Sec.~\ref{Introduction}, 5G--NR systems are the first to use frequency bands that range from microwave to mmWaves. Therefore, standardized models listed in \cite{3GPP2} (see also \cite{rappaport2017overview}) cover a very wide range of operating bands, and are in--fact valid up to 100 GHz. The parameters of the models in \cite{3GPP2} have the same style as for 4G, yet have a frequency dependent component to cater for
 bands ranging from 400 MHz to 100 GHz 
 (more measurements are needed to define /confirm the frequency dependence as given by the \cite{3GPP2}). The frequency dependence is defined for not just for pathloss but also for all model parameters needed for the impulse response. In effect, this work supersedes all earlier models\footnote{Naturally, the prior work of last 50 years leading up to these models was instrumental in providing guidance and re--calibrating our expectations of 5G pathloss models.}. Different to 4G and earlier generation models, 5G standardization describes \emph{four} categories of pathloss models each having LOS and NLOS components respectively. The four environment categories are: \emph{(1) Rural Macrocellular, (2) UMa, (3) UMi Street Canyon,} and \emph{(4) Indoor Hot Spot}. Instead of quoting the pathloss model for each of the above mentioned combinations, we describe the key modelling features. Taking the UMa NLOS environment as the axis of exposition, the following pathloss model is proposed in \cite{3GPP2}: 
\vspace{1pt}
\begin{equation}
    \label{5GPathlossModels1}
    \textrm{PL}_{\textrm{\textrm{3D UMa NLOS}}}^{\textrm{5G}}=
    \max\left(\textrm{PL}_{\textrm{3D UMa LOS}}^{\textrm{5G}},\textrm{PL}_{\textrm{3D UMa NLOS}}^{\textrm{5G(a)}}\right), 
    \vspace{3pt}
\end{equation}
where 
\begin{align}
\textrm{PL}_{\textrm{3D UMa LOS}}^{\textrm{5G}}=
    \begin{cases}
    \textrm{PL}_{1} & \quad 10\hspace{2pt}\textrm{m}<d<d_{\textrm{break}}\\
    \textrm{PL}_{2} & \quad d_{\textrm{break}}<d<5\hspace{2pt}\textrm{km}, 
  \end{cases}
\end{align}
with 
\begin{equation}
    \label{5GPathlossModels2}
    \textrm{PL}_{1}=28.0+22\log\left(d\right)+20\log\left(f\right), 
\end{equation}
and $\textrm{PL}_2$ is given by 
\begin{align}
\nonumber    
\textrm{PL}_{2} =  & 28.0+40\log\left(d\right)+20\log\left(f\right)\\
\label{5GPathlossModels3}
&-9\log\left[\left(d_{\textrm{break}}
    \right)^{2}+\left(h_{\textrm{t}}-h_{\textrm{r}}\right)^{2}\right].  \\[-22pt]
    &\nonumber
\end{align}
Furthermore, \vspace{3pt}
\begin{align}
    \nonumber
    \textrm{PL}_{\textrm{3D UMa NLOS}}^{\textrm{5G(a)}} = &
    13.54+39.081\log\left(d\right)+20\log\left(f\right)\\ 
    \label{5GPathlossModels4}
    &-0.6\left(h_{\textrm{r}}-1.5\right),
\end{align}
from 10 m to 5 km 3D link distance. As for standardized models in 4G systems, the BS and MS antenna heights in $h_{\textrm{t}}$ and $h_{\textrm{r}}$ are constrained and $d_{\textrm{break}}$ depends on the antenna heights. In order to show how pathloss varies with distance and frequency, we draw a comparison of the standardized pathloss models in \cite{3GPP2} across two carrier frequencies of 2 GHz and 100 GHz with the nominal parameters recommended in the standards. Figure~\ref{PathlossComparison} depicts the result of this comparison where we consider the pathloss for both LOS and NLOS. 
\begin{figure}[!t]
    \includegraphics[width=8.6cm]{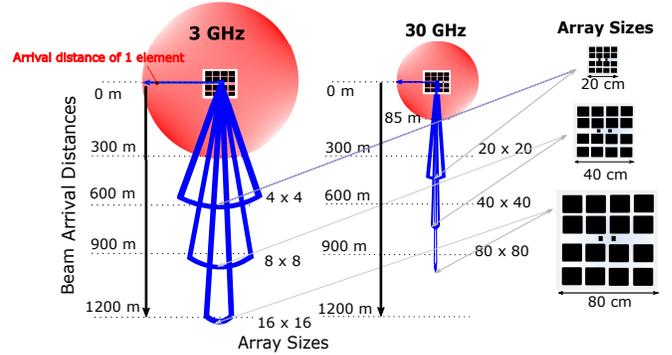}
    \hspace{-13pt}
    \vspace{-3pt}
    \caption{Demonstration of beamforming effects (in terms of beam arrival distances) relative to operating frequency and array sizes.}
    \vspace{-8pt}
    \label{BeamformingEffect}
\end{figure}
Several important trends can be observed from Fig.~\ref{PathlossComparison}. At a distance of say 100 m, the loss at 100 GHz is approximately 30 dB greater than the corresponding value for 2 GHz.  It is to be noted here that at 2 GHz, the break point distance for LOS case is around 100 m but at 100 GHz the break point distance becomes too large and is not henceforth shown on the figure. This is since the break point distance is inversely proportional to the wavelength \cite{GOLDSMITH1}. A 50 times decrease in wavelength at 100 GHz will correspondingly shift the break point 50 times further relative to 2 GHz, i.e., close to 5 km, well beyond the operating cell ranges. Moreover,   Fig.~\ref{PathlossComparison} assumes that the BS and MS antennas are \emph{omnidirectional}. However, 5G--NR deployments will be utilizing \emph{beamforming} antennas, instead of \emph{sectoral} antennas as used for earlier generation systems. Therefore, the estimation of directional pathloss is desirable. As we increase the operating frequency, higher array gain is needed to mitigate the high pathloss. An explicit illustration of this phenomenon is given in Fig.~\ref{BeamformingEffect}, which shows how the coverage range can be extended by using larger array sizes with the examples of 3 and 30 GHz carrier frequencies. 

\begin{figure}[!t]
    \vspace{-20pt}
   \hspace{-10pt}
    \centering
    \includegraphics[width=8.4cm]{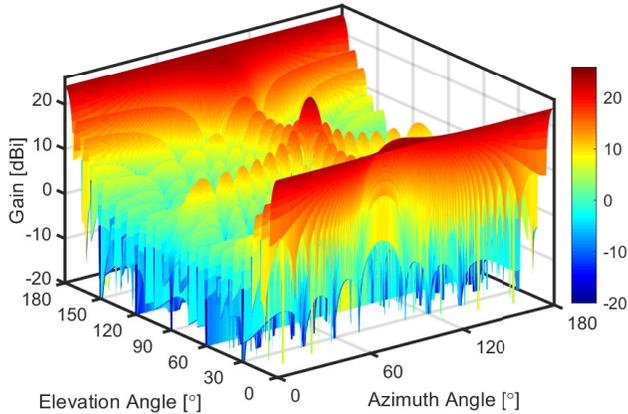}
    \vspace{-21pt}
    \caption{Directional antenna gain of a 256 element (8 (rows) $\times$ 16 (columns) $\times$ 2 (polarizations)) planar array with a horizontal inter--element spacing of 0.5$\lambda$ and vertical spacing of $0.7\lambda$, where each element has a radiation pattern which follows the description in \cite{3GPP3,3GPP2,ITU2}.}
    \label{ArrayFactor16x8x2Logical}
    \vspace{-4pt}
\end{figure}

\begin{figure}[!t] 
\vspace{-10pt}
   \hspace{-13pt}
    \centering
    \includegraphics[width=8.5cm]{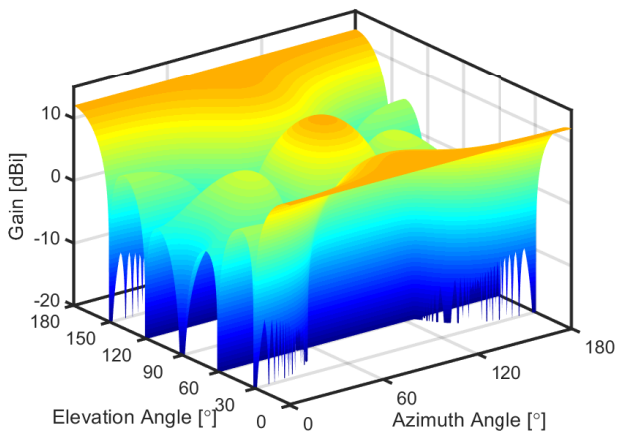}
    \vspace{-16pt}
    \caption{Directional antenna gain of a 16 element (2 (rows) $\times$ 4 (columns) $\times$ 2 (polarizations)) planar array with identical parameters as Fig.~\ref{ArrayFactor16x8x2Logical}.}
    \vspace{-19pt}
    \label{ArrayFactor4x4x2Logical}
\end{figure}

As discussed at the beginning of Sec.~\ref{pathlossmodels}, it is also possible to make the pathloss independent of frequency by keeping the physical array size at one link end independent of frequency \cite{SHAFI1,Molisch1}. Yet it is well known that as {\em electrical} array \emph{size} increases, the \emph{half--power beamwidth} (HPBW) decreases, and as a result antennas become more directional. Standardized models for 4G and 5G systems demonstrate a procedure for antenna array modelling according to a cross--polarized uniform planar array of dipole elements, having a pre--defined per--element pattern in the azimuth and zenith domains \cite{3GPP3,3GPP2,ITU2}\footnote{With the emergence of conventional and massive MIMO systems, the requirement to reduce the antenna array form factor has contributed to the stronger development of cross--polarized elements, with consideration of frequency dependent cross--polarization discrimination ratios.}. The maximum directional per--element gain is assumed to be 8 dBi and no relationship between the horizontal and vertical inter--element spacing is assumed. Naturally, the net array gain is directly proportional to the per--element pattern, number of element and the steering angle. Figure~\ref{ArrayFactor16x8x2Logical} depicts the resulting array gain as a function of the array steering angle from a 8 (rows) $\times$ 16 (columns) $\times$ 2 (polarizations) planar configuration, where a peak directional gain of 23 dBi is observed, with a HPBW of 8$^{\circ}$ in azimuth and zenith. The effect of reducing the number of elements on the directional gain is demonstrated in Fig.~\ref{ArrayFactor4x4x2Logical}, where a 2 (rows) $\times$ 4 (columns) $\times$ 2 (polarizations) configuration is employed. The peak gain is shown to reduce to 11 dBi with an increased HPBW of 63$^{\circ}$ in azimuth and 32$^{\circ}$ in zenith. For further discussions on the precise modelling methodology, the reader is referred to \cite{3GPP3,3GPP2,ITU2}. It is noteworthy that the pathloss equations presented in this section assume omnidirectional antennas, or antennas that have a constant gain over a sector. With the use of directional antennas, this assumption is violated and pathloss also becomes directional. This means that omnidirectional antenna patterns need to be synthesised from directional pathloss measurements, as done in \cite{Shusun1} with directional horn elements at both link ends. Nevertheless, majority of the directional pathloss results are very tightly linked to the channel measurement setup. A similar procedure is demonstrated in \cite{CLarsson} where an omnidirectional antenna pattern is synthesized from directional pathloss measurements. This omnidirectional pattern is then used to predict an omnidirectional pathloss. 

All of the above pathloss models are also sometimes referred to as \emph{Alpha--Beta--Gamma (ABG)} models where the ABG parameters are derived from the best fit curve match to the measured pathloss vs. distance characteristics \cite{SHAFI2}. Specifically, the Gamma parameter shows the link dependence on frequency and distance, whilst the Alpha and Beta parameters are optimised to determine an offset (intercept) ensuring the statistical fit to measured data minimizes the shadow fading standard deviation. Typically, the statistical fit is based on the least--squares method aimed to minimize the least--square chordal distance between measured and modelled curves. In this model, the offset can be fixed to the free--space pathloss at a reference distance (e.g., $1$ m); this model is also sometimes called the \emph{close--in} pathloss model \cite{Rappaport,GLOBECOMWP2}, and is used as an option (besides the ABG model) in the pathloss model for NR.

\vspace{-5pt}
\section{Standardized Impulse Response Evolution}
\label{impulseresponse}
\label{EvolutionofChannelImpulseResponses}
\subsection{Overview}
The following three sections describe the evolution of the radio channel impulse response in standardization of wireless systems. Effectively, this started from 2G and early 3G systems, where MPCs with only \emph{delays} and \emph{amplitude} parameters were considered. In such a case, the radio channel impulse response is known to be \emph{non--directional}. In contrast, impulse responses incorporating \emph{directional} channel information, such as azimuth AOD, AOA, in addition to delays and amplitudes were considered from 3G onward. At the time, this was a result of the need to increase system spectral efficiency via spatial processing, resulting in consideration of aforementioned parameters. Enhancements to 4G, such as full--dimensional MIMO (a.k.a. 3D--MIMO), required the additional consideration of zenith angles--of--departure and arrival. 5G models are impacted by
stringent requirements of the system particularly with increasing bandwidths that are possible at mmWave frequencies. 
The basic concept in the impulse response models from 3G onward is to represent a physical environment in the form of \emph{clusters} in either in geometry, or the angular domain (azimuth and zenith), determine the number of clusters, MPCs per--cluster, MPC amplitudes and phases, delays of each MPC, as well as angles--of--arrival and departure of the MPCs. To this end, the overall channel impulse response is a function of the aforementioned parameters. Below we provide a summary of non--directional and directional impulse response evolution. 

\vspace{-9pt}
\subsection{Non--Directional Models: COST 207 and ITU--R M.1225}
In 1G systems, there was no emphasis on the standardization of the methods to generate a channel impulse response, and hence no efforts were devoted to the standardization of angular and delay parameters. For the  development of the GSM system, a standardized wideband model was required to test the different system proposals. ETSI, the GSM standardization body at this time, adopted a model that had been developed by the COST 207 working group within the EU \cite{COST207}. This model defined the channel impulse response as a \emph{discrete wide sense stationary uncorrelated scattering process} for which the received signal was characterized by the summation of delayed replicas of the input signal weighted by an independent zero--mean complex Gaussian time--variant process. If $\delta(t)$, $h(t)$ denote the complex low--pass representations of the channel input and output, respectively, then: 
\vspace{-5pt}
\begin{equation}
    h\left(t,\tau\right)=\sum\limits_{n=1}^{N}\sqrt{P_{n}}\hspace{1pt}
    g_{\hspace{1pt}n}\hspace{-1pt}\left(t\right)\delta\left(t-\tau_{n}\right),
    \vspace{-5pt}
\end{equation}
where $P_{n}$ is the strength of the $n$--th MPC, and $g_{n}(t)$ is the complex Gaussian process weighting the $n$--th replica. The power spectrum of $g_{n}(t)$, known as the Doppler spectrum of the $n$--th path, controls the rate of fading due to the presence of that particular path. To completely define this model, one requires only a specification of the Doppler spectra of the MPC weights ${P}_{n}(\nu);\hspace{2pt}n=\{1,2,\dots,N\}$, the MPC delays $\tau_{n};\hspace{2pt}n=\{1,2,\dots,N\}$ and the MPC weight strengths $P_{n};\hspace{2pt}n=\{1,2,\dots,N\}$. The process $g_{n}(t)$ is to be interpreted as modelling the \emph{superposition of unresolved MPCs} arriving from different angles and in the vicinity of the delay interval satisfying 
\vspace{-1pt}
\begin{equation}
    \tau_n-\frac{1}{2B}<\tau<\tau_n+\frac{1}{2B}, 
    \vspace{3pt}
\end{equation}
where $B$ is the bandwidth of the transmitted signal. Naturally, each MPC has a different Doppler shift corresponding to a different value of the cosine of the angle between the MPC direction and its velocity vector. For the MS, the angular spectrum determines not only antenna correlation, but also the Doppler spectrum. In particular, if the MS antenna is omnidirectional, and the angular power spectrum is omnidirectional in the horizontal plane (and all MPCs arrive in the horizontal plane), then the Doppler power spectrum has the well--known ''bathtub" shape:
\vspace{-6pt}
\begin{equation}
    P\left(\nu\right)=\frac{1}{\sqrt{1-(\frac{\nu}{f_{D}})^{2}}}, 
\end{equation}
where $f_{D}=V/\lambda$ is the maximum Doppler shift, which is in turn a function of the MS speed, $V$, and carrier wavelength, $\lambda$. Note that these are the same assumptions as used by Clarke and Jakes in narrowband channel modelling \cite{Molisch1}. For certain near clusters (MPCs with lower delays), the COST 207 model also employs this modelling methodology, while for far clusters the Doppler spectrum is modeled as Gaussian. MPCs with low delay were assumed to arrive from any arbitrary direction, while long--delayed components, which arise from far scatterer clusters showed a Gaussian Doppler spectrum, corresponding to a limited spread of PAS. The COST 207 model divides radio environment was divided into 4 categories; (1) Typical urban (TU), (2) Bad urban (BU) hilly terrain (HT), (3) Rural area (RA) and (4) Hilly terrain (HT), respectively. For each of these models, the power delay profile, delay spread, and the scattering function were characterized. Generally, BU and HT show the most significant delay dispersion (with excess delays of up to 16 $\mu$s). For a precise definition of the above environment types, as well as for further details on COST 207 model, we refer the reader to \cite{COST207}.

For 3G systems, the ITU--R standardized a model known as ITU--R M.1225, was introduced in 1997. It uses the same fundamental modeling approach as the COST 207 model. 
In the interests of simplicity, the ITU--R M.1225 model makes the following assumptions: a) For outdoor channels, the model at the MS are the same as described above (leading to the bathtub spectrum). At the BS, the received MPCs arrive in a limited azimuth angular range (support). b) On the other hand, for indoor channels, a very large number of receive MPCs arrive uniformly distributed in elevation and azimuth for each delay interval at the BS. c) The antenna element is assumed to be either a short or half--wave vertical dipole. On the other hand, assumption b) results in a Doppler spectrum that is nearly flat, such that 
\vspace{-3pt}
\begin{equation}
\vspace{-2pt}
    P_{n}\left(\nu\right)=
    {2\hspace{1pt}V}; \hspace{5pt}\textrm{for}\hspace{5pt}|\nu|<\frac{V}{\lambda}.
\end{equation}
For further discussions on the ITU--R M.1225 modeling methodology, the interested reader is referred to \cite{ITUR}. In the sequel, we introduce the equivalent directional models. We first focus on the COST directional models, followed by a detailed discussion on the 3GPP/ITU--R directional models. 

\vspace{-8pt}
\section{Directionally Resolved Impulse Response Models: COST 259, 273, and 2100}
\vspace{-2pt}
\label{sec:COST_models}
During the 1990s, the concept of \emph{smart antennas} emerged as a critical tool for improving system spectral efficiency of wireless systems. The emphasis was mainly on antenna arrays (typically 4--8 elements) at the BS, allowing for either suppression of adjacent--cell interference (and thus a reduction of the spatial reuse factor/distance), or \emph{space--division multiple access}, SDMA (which would be called, in modern notation from early 2000s as multiuser MIMO following a series of pioneering information theoretic contributions of Jindal, Goldsmith, Jafar, Verdu, Caire, and Shamai, see e.g., \cite{JINDAL1,TatariaThesis} for a taxonomy). For these applications, a more detailed propagation model catering to the presence of angular dispersion (as seen by the BS) was required. The pioneering standardization activity for this was the COST 259 working group, which at the time established a first of its kind parameterized channel model for a variety of environments. Following its predecessor in COST 207, the COST 259 model \cite{Molisch259,Asplund259} equivalently considered four macrocellular environments. These are summarized as follows.  
\begin{enumerate}
\item \emph{Generalized Typical Urban (GTU):} This consists of cities where all the buildings are assumed to have uniform heights and densities. The Interacting Objects (IOs) are mostly around the MS, though there may be far scatterers. 
\item \emph{Generalized Bad Urban (GBU):} These are modern day metropolitan centers, where the building heights are not uniform with possible large open areas, parks and rivers. The MS can receive MPCs from local and far IOs. 
\item \emph{Generalized Rural Area (GRA):} This consists of farmlands, fields and forests and few buildings. The natural objects around the MS act as IOs in cities where all the buildings have uniform heights and densities. The IOs in this environment cause long detours.
\item \emph{Generalized Hilly Terrain (GHT)}: This is like the GRA, but with large height variations such as hills or mountains. Diffuse scattering from hillsides or mountains contribute significantly to the channel characteristics.
\end{enumerate}
Besides these macrocellular environments, the COST 259 model also defines microcellular environments (which consists of street canyons, intersections, and open spaces), and indoor environments (where the indoor model mainly follows the definitions of \cite{ZWICK1}); most of the subsequent discussion will relate to the macrocellular case. The COST 259 model then defines the \emph{double--directional channel impulse response} is a sum of $L$ different MPCs expressed as
\begin{equation}
\label{cost259impresp}
\underline{h}\left(\overrightarrow{r},\tau,\Omega,\Psi\right) = \sum_{l=1}^{L\left(\overrightarrow{r}\right)}
\underline{h}_{l}\left(\overrightarrow{r},\tau,\Omega,\Psi\right)
e^{j\frac{2\pi}{\lambda}\langle\overrightarrow{e}\left(\Omega_{l}\right),\overrightarrow{r}-\overrightarrow{r_{0}}\rangle}. 
\end{equation}
Note that $\underline{h}(\cdot)$ is a function of four parameters, consisting of the position vector of the MS, the delays, angles--of--arrival and angles--of--departure, respectively. As observed from \eqref{cost259impresp}, $L$ is also a function of the position vector of the MS. Furthermore, the impulse response of the $l$--th MPC is characterized by
\vspace{4pt}
\begin{equation}
\underline{h}_{l}\left(\overrightarrow{r},\tau,\Omega,\Psi\right)=\underline{\alpha_{l}}\hspace{1pt}\delta\left(\tau-\tau_{l}\right)\delta\left(\Omega-\Omega_{l}\right)\delta\left(\Psi-\Psi_{l}\right). 
\vspace{4pt}
\end{equation}
Here, $\underline{\alpha_{l}},\tau, \Omega,$ and $\Psi$ are the complex (polarimetric amplitude), delay, angles--of--departure from the BS and arrival at the MS, respectively, of the $l$--th MPC. The complex amplitude is a polarimetric 2$\times$2 matrix representing the co--polarized and cross--polarized components, respectively, such that 
\vspace{2pt}
\begin{equation}
\underline{\alpha_{l}} = \begin{bmatrix}
\alpha_{l}^{\vartheta\vartheta}&\alpha_{l}^{\vartheta\phi} \\[4pt]
\alpha_{l}^{\phi\vartheta}&\alpha_{l}^{\phi\phi}
\end{bmatrix}\hspace{-2pt},
\vspace{2pt}
\end{equation}
where $\vartheta$ and $\phi$ denote the polarizations in the H and V--planes, respectively. The phase change is computed by the location--dependent component of the arriving plane--wave, contained in the exponential factor in \eqref{cost259impresp}, 
$e^{j\frac{2\pi}{\lambda}\langle\overrightarrow{e}\left(\Omega_{l}\right),\overrightarrow{r}-\overrightarrow{r_{0}}\rangle}$. Herein, $\overrightarrow{e}(\Omega)$ denotes a unit vector pointing towards the (spatial) angle $\Omega$. From the double--directional description in \eqref{cost259impresp}, a special case of \emph{directional} channel impulse response can be obtained when multiple antennas at only one end of the link is considered with weighting of the complex polarimetric antenna radiation pattern over the transmit or receive directions. 

For outdoor macrocellular environments, the COST 259 model is in essence a geometry--based stochastic model (GSCM). It places \emph{clusters} of IOs in an area around the BS. One of those clusters is centered on the MS, while the others are placed according to a certain probability density function throughout the cell. The geographical location of the cluster center then implicitly determines the angle--of--arrival (for the uplink), as well as the MPC delays. We note that unlike the previous models, the delays and angles are implicitly correlated \cite{Molisch259,Asplund259}. For outdoor microcells, the cluster angles are determined through a quasi--deterministic modeling: BS and MS are placed on a synthetic city map (called virtual cell deployment area), and the center angles as well as delays of the clusters are determined by tracing the dominant paths ``down the street". Due to the geometric placement of the clusters in the COST family of models, the changes of the cluster angles and MPC delays as the MS moves are \emph{implicitly} modelled.

\begin{figure}[!t]
    \centering
    \includegraphics[width=8.5cm]{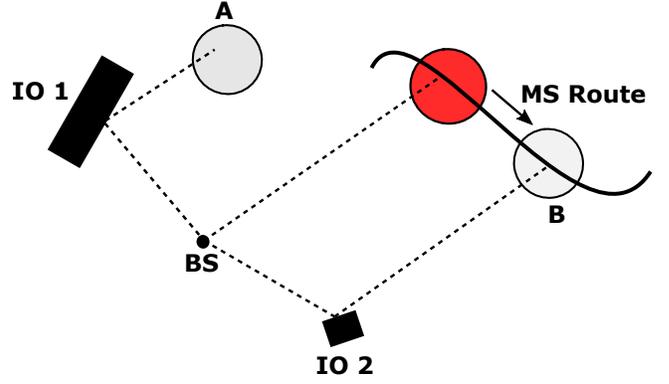}
    \caption{Example of visibility areas (gray circles) and IO positions (black rectangles). The red circle is the local scattering cluster, which moves with the MS. Along the particular MS trajectory shown, only one of the two clusters (B) will be activated.}
    \vspace{-10pt}
    \label{COST259Clusters}
\end{figure}

Furthermore, COST 259 introduced the concept of {\em visibility region (VR)} for the channel modeling. In essence, each cluster is associated with one or more, randomly distributed (according to a certain probability distribution function) region, and the contributions of a cluster to the impulse response are only nonzero if the MS is in the VR of that cluster.
This concept is depicted in Fig.~\ref{COST259Clusters}, where the visibility areas of different clusters A and B (indicated with gray circles) are presented. Here, as the MS moves, one of the two clusters shown will be activated, with the red circle denoting the local scattering also moving with the MS. Further discussions on spatial consistency is presented later in the paper. 
The geometric computation of the cluster location and the VRs 
guarantee \emph{spatial consistency} in the channel description. The above processes determine the parameters of the \emph{cluster centers}.  The cluster dispersion with respect to delay and angle at the BS is modelled as separable, such that the azimuthal delay power profile (ADPD) can be written as
\vspace{-2pt}
\begin{align}
\nonumber
P_{\tau,\phi,i}(\tau,\phi)&=P_{\tau,i}\hspace{1pt}(\tau) P_{\phi,i}\hspace{1pt}(\phi)\\[8pt]
&=C\left[\hspace{1pt}e^{(-\tau/S_{\tau,i})}\hspace{1pt}
e^{(-|\phi|/S_{\phi,i}\hspace{1pt})}\right], \\[-18pt]
\nonumber
\end{align}
where $i$ is the cluster index, $C$ is a constant of proportionality, $S_{\tau,i}$ is the cluster delay spread and $S_{\phi,i}$ angular spread, respectively. This model drew from the celebrated \emph{Saleh--Valenzuela} model \cite{Saleh1}, in that it describes the cluster power delay profile as exponentially decaying, while modeling the azimuthal spread as Laplacian, based on extensive measurement campaigns (see e.g., \cite{Algans_et_al_1999}). The model describes also the angular dispersion at the MS. However, only angular dispersion at {\em either} BS {\em or} MS can be simulated in general, since angles at the two link ends are correlated through the position of the scatterers; this limitation was lifted in the COST 273 model (see below). 
Importantly, the COST 259 model provides a \emph{continuous} ADPD; in order to obtain a discrete simulation model that is tailored to a particular bandwidth, a suitable discretization must be developed; yet this is not part of the actual standardized model--we will see that this is in contrast to the later 3GPP/ITU--R models. The model was then parameterized based on measurements both in the literature, and those done within COST 259 \cite{Molisch259},\cite{Asplund259}. 

The models of COST 273 and COST 2100 are largely built on the methodology of COST 259, but in a more generalized manner to incorporate angular dispersion at \emph{both} link ends. In this context, it is important to distinguish between MPCs that interact with other IOs only \emph{once} (single scattering), and those where \emph{multiple} interactions occur (multiple scattering). In the former case, the directions at the BS and at the MS are linked, and the above description for cluster placement can be used for MIMO as well. In the latter case, the directions at the BS are determined by the location of the IO closest to the BS that the particular MPC interacts with, while the directions at the MS are determined by the IO (for this MPC) closest to the MS; in other words, \emph{directions at both the BS and MS become spatially decoupled and independent from the delays}. The COST 273 model thus introduces the concept of ``twin clusters" that determine the directions and (probabilistically determines) the excess delay that describes the propagation between those clusters \cite{Hofstetter_and_Molisch_2006}. The COST 2100 model further refined these ideas and provided parameterizations in a variety of different environments \cite{COST2100}. It also introduced the concept of \emph{joint clusters} to model correlations between MSs in multiuser scenarios in a more realistic way \cite{Poutanen_et_al_2012}. In later COST actions, such as COST Information Communications 1004 (IC 1004) and COST Inclusive Radio Communications (IRACON), the COST 2100 model was also extended for massive MIMO channels by modelling non--stationarities over a large antenna array, and death--birth processes for individual MPCs \cite{Flordelis20_et_al}.

\begin{table*}[!t]
\centering
\scalebox{0.485}{
\begin{tabular}{l ccc ccc ccc ccc}
\toprule
\midrule
 & \multicolumn{3}{c}{\textbf{UMi--Street Canyon}} & \multicolumn{3}{c}{\textbf{UMa}} & \multicolumn{3}{c}{\textbf{RMa (Upto 7 GHz)}} & 
 \multicolumn{3}{c}{\textbf{InH}}\\
 \midrule
    & \textbf{LOS}   & \textbf{NLOS}  & \textbf{O2I} & \textbf{LOS}   & \textbf{NLOS}  & \textbf{O2I}  & 
    \textbf{LOS} & \textbf{NLOS} & \textbf{O2I} & 
    \textbf{LOS} & \textbf{NLOS} & \\ 
\textbf{AOD Spread (ASD)} &$\mu_{\textrm{lg,ASD}}=-0.05\log_{10}\left(1+f_{c}\right)+1.21$ & $-0.23\log_{10}(1+ f_{c}) + 1.53$ & $1.25$& 
$1.06+0.1114\log10(f_{c})$& $1.5-0.1144\log_{10}(f_{c})$ & 1.25 & 0.90 & 0.95 & 0.67 & 1.60 & 1.62\\ &$\sigma_{\textrm{lg,ASD}}=0.41$ & $0.11\log10(1+ f_{c})+0.33$& 0.42& 0.28 & 0.28 & 0.42 & 0.38 & 0.45 & 0.18 & 0.18 & 0.25 \\ 
\midrule
\textbf{AOA Spread (ASA)} & $\mu_{\textrm{lg,ASA}}=-0.08\log_{10}(1+ f_{c})+1.73$  & $-0.08\log_{10}(1+ f_{c})+1.81$ & 1.76 & 1.81 & $2.08-0.27\log_{10}(f_{c})$ & 1.76 & 1.52 & 1.52 & 1.66 & $-0.19 \log_{10}(1+f_{c}) + 1.781$ & $-0.11\log_{10}(1+f_{c})+1.863$\\
&$\sigma_{\textrm{lg,ASA}}=0.014\log_{10}(1+ f_{c})+0.28$ & $0.05\log_{10}(1+ f_{c})+ 0.3$& 0.16 & 0.20& 0.11&0.16&0.24&0.13&0.21& $0.12\log_{10}(1+f_{c})+0.119$ & $0.12 \log_{10}(1+f_{c})+0.059$ \\ \midrule
\textbf{ZOA Spread (ZSA)} &  $\mu_{\textrm{lg,ZSA}}=-0.1\log_{10}(1+ f_{c})+0.73$& $0.05 \log_{10}(1+ f_{c})+0.3$ & 0.16 & 0.20 & 0.11 & 0.16 & 0.47 & 0.58 & 0.93 & $-0.26\log_{10}(1+f_{c})+1.44$ & $-0.15\log_{10}(1+f_{c}) + 1.387$ \\ 
& $\sigma_{\textrm{lg,ZSA}}=-0.04\log_{10}(1+f_{c})+0.34$ 
& $-0.07\log_{10}(1+ f_{c})+0.41$& 0.43&0.16&0.16&0.43 & 0.40 & 0.37 & 0.22 & $-0.04\log_{10}(1+f_{c})+0.264$ & $-0.09\log_{10}(1+f_{c})+0.746$\\ \midrule
\textbf{Delay Spread (DS)}& $\mu_{\textrm{lg,DS}}=-0.24\log_{10}(1+ f_{c})-7.14$ & $-0.24 \log_{10}(1+ f_{c})-6.83$ & -6.62& $-6.955-0.0963\log_{10}(f_{c})$ &$-6.28-0.204\log_{10}(f_{c})$ &-6.62&-7.49 & -7.43 &-7.47 & $-0.01\log_{10}(1+f_{c})-7.692$& $-0.28\log_{10}(1+f_{c})-7.173$\\
& $\sigma_{\textrm{lg,DS}}=0.38$ & $0.16\log_{10}(1+f_{c})+0.28$ &0.32 &0.66 & 0.39& 0.32&0.55&0.48&0.24&0.18& $0.10\log_{10}(1+f_{c})+0.055$\\
\midrule
\bottomrule
\end{tabular}}
\vspace{5pt}
\caption{Lognormal angular and delay spread mean and standard deviation parameters for the various different environments defined by 3GPP. The defined parameters are categorized for LOS, NLOS and O2I states, respectively.}
\label{AngularandDelayParameters}
\vspace{-15pt}
\end{table*}

\vspace{-9pt}
\section{Directionally Resolved Impulse Response Models: 3GPP/ITU--R Models}
\label{sec:3GPP}
\subsection{Motivation}
While the COST models are highly refined and detailed, they were rather complicated for system level simulations in the development of standardized 3G (and later 4G, as well as 5G) wireless systems. Furthermore, the absence of a discretized version of the COST 259 model that could be used by all standards participants was deemed a stumbling stone. Thus, the rivalling standardization bodies 3GPP and 3GPP2 teamed up to develop a channel model that could be used for comparisons of different systems proposals \cite{3GPP25996,CALCEV1}. In parallel to this, the ITU--R also developed their own models for 4G and 5G systems, which are aligned with the those proposed by the 3GPP. Collectively, these models are usually referred to as the 3GPP and ITU--R models, and are based on many of the concepts presented in the COST models, yet also shows some important differences. In particular, even though the 3GPP/ITU--R model standards describe the model as a GSCM, the implementation of the angular dispersion can actually be better described as extension of a tapped--delay--line model to the angular domain. Below we describe the 3GPP/ITU--R model structure and explain the parameters. 
\vspace{-10pt}
\subsection{Model Structure}
\label{structure}
The 3GPP model consists of a number of \emph{paths}, each of which has a  particular delay (in later versions, the paths are also called ``clusters", but are different from the earlier cluster definition in COST, as they do not exhibit delay dispersion). The delays of those paths is either given deterministically in one form of the model, or are determined at random, according to a given (parameterized) probability density function. Power is assigned according to the path delay, with the average power (over shadowing) decreasing with increasing delay. Turning now to the angular dispersion: the ``baseline" of the angles at both BS and MS is the LOS connection between the paths (such a connection can be drawn even for those cases that an actual physical LOS component does not exist). Each of the paths has a deviation from the baseline (equally likely to the right and the left from the LOS), which is created according to a specified probability density; and which increases with increasing delay of the path. Last but not least, each path (or cluster) has itself an angular spread, which takes on a deterministic value, such as 5 degrees. This is realized in that a cluster consists of 20 sub--paths, which all have the same delay, but slightly different angles. Each of the sub--paths has the same amplitude, and random phases; their superposition thus provides not only an angular spread, but also small--scale fading when either the MS moves, or different values of the random phases are chosen. Note that the angle deviations at BS and MS are chosen independently (i.e., no pure single scattering is modeled), but they are still somewhat correlated in that large delays lead to large deviations from the baseline angle at \emph{both} link ends. As for the COST model, the 3GPP/ITU--R models allow for the directional impulse response to be evaluated with cross--polarized antenna elements across both H and V--polarizations. The evolution of the model to 4G and 5G systems is presented further in the text. 

Using the same notation as for 3GPP TR 25.996, the channel coefficients for each cluster $n$ and each MS and BS element pair, $(u,s)$, is given by \cite{3GPP3,ITURM2135,3GPP2,ITU2} 
\vspace{-2pt}
\begin{align}
    \nonumber
    h_{u,s,n}\left(t\right)=\hspace{3pt}&
    \sqrt{\frac{P_{n}}{M}}\sum\limits_{m=1}^{M}
    \begin{bmatrix}
        F_{\textrm{MS},u,\theta}\left(\theta_{n,m,\textrm{ZOA}},\phi_{n,m,\textrm{AOA}}\right)\\
       F_{\textrm{MS},u,\phi}\left(\theta_{n,m,\textrm{ZOA}},\phi_{n,m,\textrm{AOA}}\right)
    \end{bmatrix}^{T}\\ \nonumber
    &\hspace{-45pt}\times\begin{bmatrix}
    \textrm{exp}\left(j\Phi_{n,m}^{\theta,\theta}\right) & 
    \sqrt{\kappa_{n,m}^{-1}}\hspace{2pt}\textrm{exp}\left(j\Phi_{n,m}^{\theta,\phi}\right)\\ 
    \sqrt{\kappa_{n,m}^{-1}}\hspace{2pt}\textrm{exp}\left(j\Phi_{n,m}^{\phi,\theta}\right) & 
    \textrm{exp}\left(j\Phi_{n,m}^{\phi,\phi}\right)
    \end{bmatrix}\\[5pt] \nonumber
    &\hspace{-45pt}\times\begin{bmatrix}
        F_{\textrm{BS},s,\theta}\left(\theta_{n,m,\textrm{ZOD}},\phi_{n,m,\textrm{AOD}}\right)\\
       F_{\textrm{BS},s,\phi}\left(\theta_{n,m,\textrm{ZOD}},\phi_{n,m,\textrm{AOD}}\right)
    \end{bmatrix}
    \hspace{-2pt}\textrm{exp}\hspace{-2pt}\left(\hspace{-1pt}j2k\hspace{-1pt}\hspace{-1pt}\left(\hspace{-1pt}\overrightarrow{r}_{\hspace{-3pt}\textrm{MS},n,m}^{T}\hspace{-3pt}\overrightarrow{d}_{
\hspace{-2pt}\textrm{MS},u}\right)\hspace{-1pt}
    \right)\\[5pt]
    &\hspace{-45pt}\times\textrm{exp}\hspace{-1pt}\left(j2k
    \hspace{-1pt}\left(\overrightarrow{r}_{\textrm{BS},n,m}^{T}\hspace{-1pt}\overrightarrow{d}_{\hspace{-2pt}\textrm{BS},s}\right)\right)
    \textrm{exp}\left(j2\pi{}v_{n,m}t\right). \\[-17pt]
    &\nonumber
\end{align}
Here, $k=2\pi\lambda^{-1}$ is the wave number relative to the carrier frequency, $F_{\textrm{MS},u,\theta}$ and 
$F_{\textrm{MS},u,\phi}$ are the $u$--th receive antenna element radiation patterns in the direction of the spherical basis vectors, 
$\overrightarrow{\theta}$ and $\overrightarrow{\phi}$, respectively.  Also, $F_{\textrm{BS},s,\theta}$ and 
$F_{\textrm{BS},s,\phi}$ are the $s$--th transmit antenna element  radiation patterns in the direction of the spherical--F
basis vectors, $\overrightarrow{\theta}$ and $\overrightarrow{\phi}$, respectively. Furthermore, $\overrightarrow{r}_{\textrm{MS},n,m}$ is the spherical unit vector with azimuth arrival angle $\phi_{n,m,\textrm{AOA}}$ and elevation arrival angle $\theta_{n,m,\textrm{ZOA}}$, given by 
\begin{equation}
\overrightarrow{r}_{\textrm{MS},n,m}=
    \begin{bmatrix}
    \sin\left(\theta_{n,m,\textrm{ZOA}}\right)
    \cos\left(\phi_{n,m,\textrm{AOA}}\right)\\
      \sin\left(\theta_{n,m,\textrm{ZOA}}\right)
    \sin\left(\phi_{n,m,\textrm{AOA}}\right)\\
    \cos\left(\theta_{n,m,\textrm{ZOA}}\right)
    \end{bmatrix},
\end{equation}
where $n$ denotes a cluster and $m$ denotes a ray within cluster $n$. Note that $\overrightarrow{r}_{\textrm{BS},n,m}$ is the spherical unit vector with azimuth departure angle $\phi_{n,m,\textrm{AOD}}$ and elevation departure angle $\theta_{n,m,\textrm{ZOD}}$, given by 
\begin{equation}
\overrightarrow{r}_{\textrm{BS},n,m}=
    \begin{bmatrix}
    \sin\left(\theta_{n,m,\textrm{ZOD}}\right)
    \cos\left(\phi_{n,m,\textrm{AOD}}\right)\\
      \sin\left(\theta_{n,m,\textrm{ZOD}}\right)
    \sin\left(\phi_{n,m,\textrm{AOD}}\right)\\
    \cos\left(\theta_{n,m,\textrm{ZOD}}\right)
    \end{bmatrix}. 
    \vspace{3pt}
\end{equation}
Further to this, $\overrightarrow{d}_{\textrm{MS},u}$ is the location 
vector of the receive antenna element $u$. Similarly, 
$\overrightarrow{d}_{\textrm{BS},s}$ is the location vector of transmit element $s$ and $\kappa_{n,m}$ is the cross--polarization power ratio. If uni--polarized antennas are assumed, then the $2\times{}2$ polarization matrix can be replaced by $\textrm{exp}(j\Phi_{n,m})$. It is to be noted that the Doppler frequency component $v_{n,m}$ is calculated from the arrival angles (AOA, ZOA), MS velocity vector, $\overrightarrow{v}$ with
speed $v$, travel azimuth angle $\phi{v}$, and elevation angle $\theta{}v$. It is given by 
\begin{equation}
    v_{n,m}=\frac{\overrightarrow{r}_{\textrm{BS},n,m}^{T}\hspace{2pt}
    \overrightarrow{v}}{\lambda_{0}^{-1}},
\end{equation}
where $\overrightarrow{v}=v[\sin(\theta_v)\cos(\phi_v)\hspace{7pt}
\sin(\theta_v)\cos(\theta_v)\hspace{7pt}\cos(\theta_v)]^{T}$. 
\begin{figure*}[!t]
\vspace{-5pt}
    \centering
    \includegraphics[width=16.5cm]{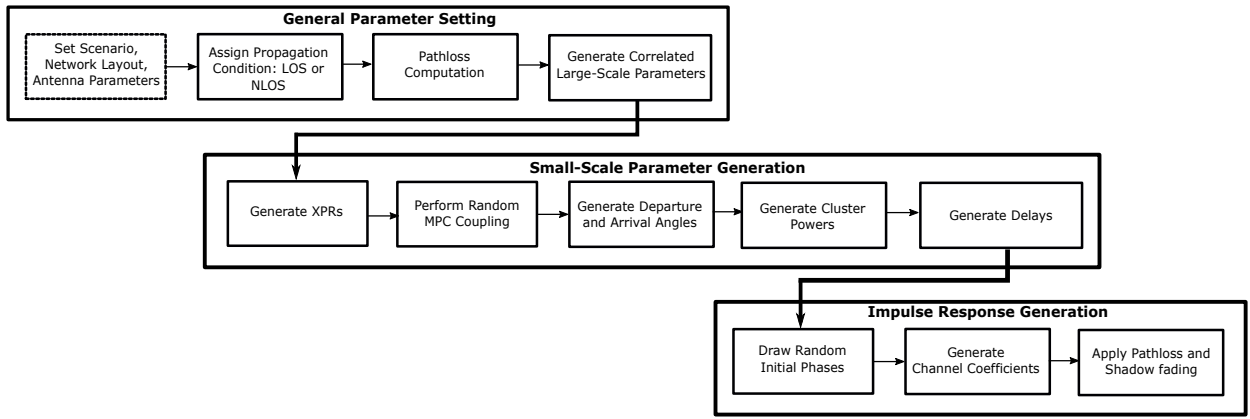}
    \caption{Standardized 3GPP/ITU--R channel impulse response generation procedure for propagation channel models in 4G and 5G systems.}
    \label{impulseresponsegeneration}
    \vspace{-12pt}
\end{figure*}

\vspace{-5pt}
\subsection{Parameterization for Different Environments}
The angular spreads in the different environments can be usually described as random variables that are lognormally distributed, so that a description of mean and standard deviation is sufficient. These values are prescribed in the standard separately for LOS, NLOS and (in applicable cases) for Outdoor--to--Indoor (O2I) situations. The values are provided in Table \ref{AngularandDelayParameters}. For the four environments of UMi--street canyon, UMa, RMa (up to 7 GHz) and InH, one can notice a frequency dependency of the AOD, AOA, ZOA and delay spread mean and standard deviations. This is since the target of 3GPP 38.901/ITU--R M.2412 models was to support 3GPP Release 15 for 5G--NR systems. The exact relationship between the quoted values across all the parameters is a complex task, one which is best evaluated via detailed propagation measurements. 
\vspace{-9pt}
\subsection{Historical Evolution}
\vspace{2pt}
The original 3GPP model, TR 25.996, was a simplification of the COST 259 model. Yet, relative to the ITU--R M.1225 model, a phenomenal improvement was made in this model, as it included the small--scale \emph{angular} parameters, and the model parameters included the instantaneous path gain and phase, shadow fading, azimuth AOA and AOD, directional dependent BS and MS antenna gains, polarization parameters, as well as magnitude and direction of the MS velocity vector. This was also the first model where support for multiple antenna capability at the BS and MS link ends was provided with specific pre--defined antenna array parameters. On the other hand, TR 25.996 still had significant gaps. It was only defined for the suburban macrocellular, urban macrocellular, and urban microcellular. Over the years, parameterization was extended to a much larger range of environments, mainly based on measurements of the European Union WINNER and WINNER II projects. The parameters such as delay spread and angular spread were extracted from measurements by a variety of consortium partners, using various propagation channel sounders. Furthermore, elevation parameters were included, and hence provided the ability to model MPCs departing/arriving from clusters of scatterers in terms of amplitudes, phases, delays, azimuth and elevation angles and polarization parameters. This model was extensively used for investigations into full--dimensional MIMO. All these additions resulted in 3GPP TR 36.873, and were adopted by the ITU--R as model M.2135. These are the models that were used for the standardization of Long--Term Evolution (LTE) and IMT--Advanced systems. Attempts were made to further generalize the models to larger frequency ranges, in particular mmWave channels, for 5G cellular systems. A Special Interest Group (SIG) made extensive proposals \cite{Haneda_et_al_2017ab}; \emph{However, the ultimately adopted specifications for 38.901 show little dependence on the carrier frequency (apart from the parameters reported in Table~\ref{AngularandDelayParameters}). Thus, while claiming to be valid up to 100 GHz, the model mainly has an experimental basis only for $<6$~GHz.} Further modifications in the model are related to  large bandwidths, large antenna arrays (i.e., for massive MIMO), and spatial consistency for mobility simulations.

\vspace{-9pt}
\subsection{Step Wise Procedure for Impulse  Response Generation}
A step wise procedure for the generation of impulse response is shown in Fig.~\ref{impulseresponsegeneration}. In the interest of brevity, this is now briefly described but details can be found in \cite{3GPP25996,3GPP2,3GPP3}. Due to the evolution of complexity, there are slight differences in the impulse response generation in \cite{3GPP25996} and\cite{3GPP2,3GPP3}. First, one can choose a network scenario, e.g. rural macrocellular, urban macrocellular, urban microcellular, etc., as well as the associated radio system layout, antenna parameters (numbers of elements, antenna gain, and beamforming architecture).  Then the pathloss is computed for the MSs in LOS and NLOS conditions, followed by correlated large--scale propagation parameters. At this stage, the generation of small-scale parameters begins, involving computation of the angles--of--arrival and departure in azimuth and zenith for each user given its location, the random delays, cluster powers, cross--polarization discrimination values. All of these parameters are random variables with specified distributions. This is then applied to the relevant expressions in \cite{3GPP3,ITU3,3GPP2,ITU2} to generate the impulse response. This process is repeated for all the MSs via the principles of a ``drop" where each drop randomly initializes different MS position.

\vspace{-8pt}
\section{Site--Specific Channel Modeling} 
The need of radio channel modeling that incorporates geometrical information of cellular site is essential to relate multi--dimensional multipath channel parameters, such as locations of communication devices and scatterers and propagation delays and angles to each other, as discussed in the previous sections. They have been so far discussed in the context of stochastic channel models in this paper, but the geometrical information is also essential in site--specific channel modeling. The need of such modeling comes mainly from {\it coverage planning} where geometry of cellular sites influences the feasible BS configurations. (Quasi)--deterministic modeling of channels are required for this purpose. Such modeling of channels are naturally applicable to any radio frequencies as far as mathematical models of relevant wave propagation phenomena are available. 

This section provides an overview of site--specific channel modeling methods. They have evolved from the use of a simplified geometry and wave propagation mechanisms so that they can run with moderate computational load and decent channel prediction accuracy, to state--of--the--art ones incorporating accurate geometries along with elaborated propagation mechanisms and empowered by high performance computational devices. Finally, the use of site--specificity in stochastic models to improve spatial consistency is discussed. 

\vspace{-10pt}
\subsection{Seminal Site--Specific Models}
The first site--specific channel models are based on ray--based channel modeling in an environment with extremely simplified geometry. This is because very accurate geometrical databases of built--up environments were not available back then. The papers of Ikegami, Walfisch and Bertoni published in 1980s for pre--2G and 2G systems elaborated approaches to estimate propagation pathloss in urban cellular environments where rooftops of office buildings and residential houses are modeled as a series of absorbing screens, e.g.,~\cite{Ikegami84TAP,WALFISCH1} and \cite{Bertonibook}, Chapter 6. Such a simplified approach was complemented by experimental correction terms for better reproduction of measured pathloss, and furthermore, applied to different parts of urban environments, e.g., street intersections~\cite{COST231}, Chapter 4.4. The ray--based channel modeling in a simple city geometry later evolved into reference channel modeling introduced in the previous sections, driven by the demand to compare candidate physical layer technologies for standardization. A series of European Initiatives made significant contributions to the evolution. In addition to them, more accurate site--specific channel modeling has become feasible thanks to improvement in available computational power and mathematical modeling of wave propagation phenomena.

\vspace{-10pt}
\subsection{Ray--Based Wave Propagation Simulation Techniques}
Over the past years, channel modeling based on geometrical information of the cellular site has been more feasible and attractive because of the increase in available computational power and ray calculation methods taking advantage of parallel computing. Computationally efficient ray--launching algorithms~\cite{Lu19TAP,He19CST} are supported by graphical processing unit or clustered computers, parallelized identification of ray optical paths over the geometry and adaptive density of launched rays. The computational efficiency allows cellular coverage study in very wide areas such as macrocells in San Francisco~\cite{Lu19TAP}. Ray--tracers are also widely used to study channels where channel sounding is not easily possible, e.g., in drone and high--speed railway scenarios~\cite{Yang19TVT}.

\vspace{-8pt}
\subsection{Improved Models for Wave Scattering and Link Shadowing}
Increasingly powerful ray--tracing implementation also allows incorporating more complex propagation mechanisms than reflection and diffraction. Diffuse scattering due to electrically rough surfaces is one of such propagation mechanisms that contribute to link gains. Various statistical models of scattering~\cite{Degli-Esposti07TAP} support its implementation into quasi--deterministic channel modeling. While its contribution to link gains naturally differs for environments and frequencies, recent studies show their impacts in outdoor mmWave channels~\cite{Mani18EuCAP}. 

Link shadowing is another example of wave--object interaction that requires large computational efforts to be properly considered in quasi--deterministic channel modeling. Vegetation for example causes link shadowing as well as wave scattering in outdoor scenarios. Their mathematical models are studied at various below--and above--6 GHz frequencies~\cite{Mani12TAP,Leonor19TAP,Torrico19RS}, all of which reproduce the measured reality. Human bodies can also cause link shadowing, especially for above--6~GHz frequencies. Simple physically-motivated models to estimate link blockage loss due to a human body calculates diffracted fields from various shapes of blocking objects~\cite{Ghaddar07TAP,Kunisch08ICUWB,Virk19TAP}. As many physical objects become electrically large as the carrier frequency is higher, their inclusion into quasi--deterministic channel modeling becomes more essential for a good site--specific coverage study. 

\vspace{-12pt}
\subsection{The Use of Point Cloud Models of the Environment}
\begin{figure}[!t]
\vspace{-10pt}
\centerline{\includegraphics[width=0.9\columnwidth,draft=false]
    {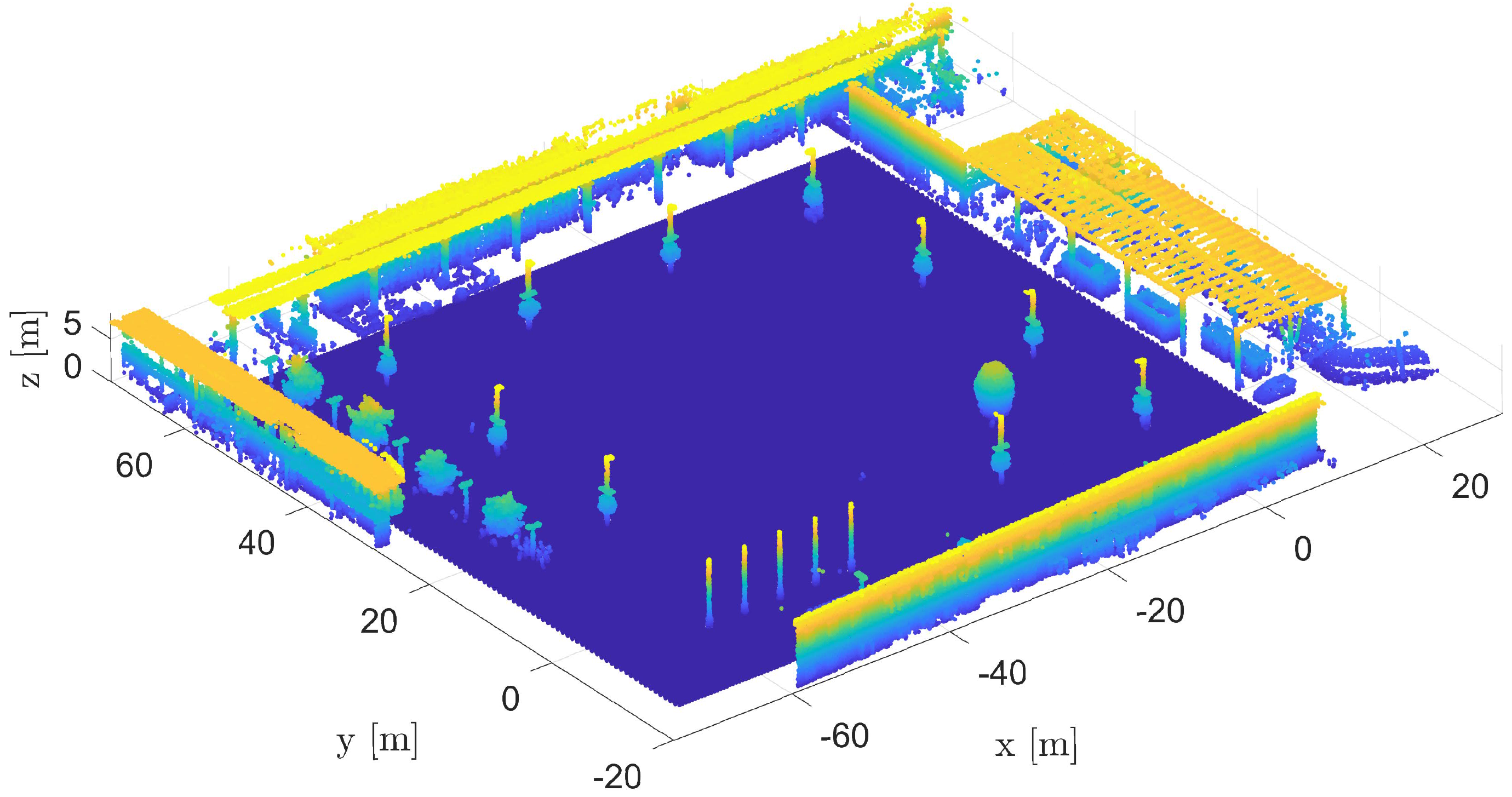}}
\caption{A sample point cloud of an open square. Points are colored according to their heights above the ground.} 
\label{fig:point_cloud}
\vspace{-14pt}
\end{figure}

The use of point clouds for quasi--deterministic channel modeling has been discussed to aim at more accurate site--specific channel modeling~\cite{Jarvelainen16TAP,Inomata18EuCAP,Stephan18EuCAP}. Point clouds are obtained by optical measurements of a site, e.g., by cameras and laser. They include detailed structures of the site, including lampposts, trees and facade structures in outdoor scenarios, and tables, chairs and signboards indoors, which are not usually documented in commercial digital maps. An exemplary point cloud from an open square is illustrated in Fig.~\ref{fig:point_cloud}. The use of point cloud is advantageous at higher frequencies as the physically small details becomes electrically large, leading to possible noticeable effects of those details on coverage. The work~\cite{Koivumaki20TAP} shows that point cloud--based modeling simulates multipaths in a more consistent manner to measurements.

\vspace{-13pt}
\subsection{Extension to Reference Channel Modeling}
\vspace{-3pt}
Methods in site--specific channel modeling have helped improvement of stochastic channel modeling. As discussed in the previous sections, spatial consistency of the channels is implicit in site--specific channel model as communication devices and wave scatterers are defined by their coordinate systems of a cellular environment. Stochastic channel models can therefore take advantage of the knowledge of geometry to ensure spatial consistency. For example, \cite{Molisch02VTCS} defines a virtual cell deployment area, similar to the Manhattan grid, as a part of the COST 259 microcellular channel model~\cite{COST259book}. The virtual deployment area restricts possible scatterer locations to physically meaningful places such as walls, corners and rooftops of buildings. Defining locations of communication devices and clusters on a virtual area allows explicit relation between delay and angles of multipath components, naturally leading to spatial consistency. The same idea has later been implemented in the European Mobile and Wireless Communications Enablers for the Twenty--twenty Information Society (METIS) project~\cite{MEDBO1}, showing good comparison with measured pathloss. Similarly, statistical generation of clusters on physically meaningful places on a geometry is useful for non--cellular scenarios, as proved by the models for vehicular communications, where clusters are restricted mostly to locations of cars and roadside objects in high way scenarios~\cite{Karedal09TWC}, or along walls in urban intersections \cite{Gustafson20TWC}.

\vspace{-13pt}
\subsection{Map--Based Channel Models for 5G Systems}
Map--based hybrid channel models are also described in \cite{3GPPTR38900} for 5G systems, though the procedure described in this reference is limited to $0.5-6$~GHz. The radio channels are created using ray tracing  on a digital map. The digitized map should contain 3D geometric information of all major structures like buildings, construction materials, random small objects in microcells etc. Network layout and antenna parameters are then set. Finally the clusters are deterministically placed. From here onward, the impulse response generation is similar to what is described earlier in Sec.~\ref{sec:3GPP}.
\vspace{-8pt}
\section{Additional Modelling Components}
\subsection{LOS Probability Modelling}
The LOS probability denotes the probability that a MS experiences geometric LOS propagation conditions with respect to the BS location\footnote{We clarify that according to its standardized definition in 3GPP TR 25.996, the LOS component refers to a geometric LOS component with no per--path angular spread \cite{3GPP25996}.}. LOS probability started to feature in standardized propagation models from 3G (COST 259 and 3GPP TR 25.996 \cite{3GPP25996} onward), since 1G and 2G systems primarily focused on the modelling of amplitude and delay parameters for the considered environments. In standardized models for 3G, 4G and 5G systems \cite{3GPP25996,3GPP2,3GPP3}, the LOS \emph{state} is determined by considering the obstruction (interruption) of the geometric LOS path between the BS and MS. It is important to note that the impact of the other IOs, such as trees, cars, buildings is not catered for in the 3GPP  methodology, and is usually modelled separately via additional shadowing/blockage terms. Rather interestingly, since IO details in the propagation channel are not taken into account, \emph{the LOS probability is not considered to be a function of the carrier frequency}. This is also the case for the multi--organization white paper channel model (5GCM) presented in \cite{GLOBECOMWP1}.\footnote{On the other hand, if frequency dependent artifacts are taken into account in the modelling of the LOS probability, it indeed would be a function of the carrier frequency. As shown in Fig.~A.4.1.2-1(b) of \cite{GLOBECOMWP2}, inclusion of the above yields a direct correlation between the LOS probability and carrier frequency, as well as relative heights of the BS and MS.} In general, standardized LOS probability models have been developed for both indoor and outdoor UMi and UMa scenarios. The UMi scenarios include high MS density open areas and street canyons with below rooftop BS heights (e.g., 3--20 m) and nominal MS heights at the ground level (around 1.5 m). Inter--site distances (ISDs) of 200 m or less has been discussed in \cite{3GPP3}. In contrast to this, the UMa scenarios typically have BSs mounted above rooftops of surrounding buildings (e.g., 25–-30 m) with MS heights around 1.5 m and ISDs of up to 500 m. In what follows, we provide a synopsis of the standardized models for probability of LOS in the aforementioned environments. 
\subsubsection{3GPP TR 36.873/38.901 UMi LOS Probability \cite{3GPP2,3GPP3}}
This model is often referred to as the $d_1/d_2$ model, where the probability of LOS is characterized by 
\begin{equation}
    \label{PLOS3GPP}
    P\hspace{1pt}_{\textrm{LOS}}^{\textrm{UMi}}
    \left(d\hspace{1pt}\right)=\min
    \left(\frac{d_1}{d},1\right)
    \left(1-e^{-d/d_2}\right)+e^{-d/d_2}, 
\end{equation}
where $d$ is the 2D link distance from the BS to the MS, $d_1$ and $d_2$ are reference distances optimized to fit a set of scenario parameters, as described in \cite{3GPP2,3GPP3}. The model parameters were found to be
$d_1=18$m and $d_2=36$m \cite{3GPP1,3GPP2}. For a link between
an \emph{outdoor BS} and an \emph{indoor MS}, the model uses the outdoor distance, $d_{\textrm{2D-out}}$, which is the 2D/ground distance from the BS to the surface of the indoor building. 
\subsubsection{3GPP TR 36.873/38.901 UMa LOS Probability \cite{3GPP2,3GPP3}}
This model expresses the LOS probability as a function of distance by 
\vspace{-3pt}
\begin{align}
\nonumber
 P\hspace{1pt}_{\textrm{LOS}}^{\textrm{UMa}}\left(d\hspace{1pt}\right)&=\left[\hspace{1pt}\min\left(\frac{d_1}{d},1\right)\left(1-e^{-d/d_2}\right)+e^{-d/d_2}\right]\\[5pt]
 &\times\Big[\hspace{1pt}1+C\left(d,h_{r}\right)\Big], \\[-24pt]
 &\nonumber
\end{align}
where 
\begin{align}
    C\left(d,h_{r}\right)=
    \begin{cases}
    0 & \quad \hspace{2pt}h_{\textrm{r}}<13\textrm{m}\\
    \left(\frac{h_{\textrm{r}}-13}{10}\right)^{1.5}g\left(d\right)& \quad 13\textrm{m}\leq{}h_{\textrm{r}}\leq{}23\textrm{m}, 
  \end{cases}
\end{align}
with $g(d)=\left(1.25\hspace{1pt}e^{-6}\right)d^{\hspace{1pt}2}e^{-d/150}$, if $d>18$ or $0$ otherwise. Here the  antenna heights at the BS and MS are readily visible in the model unlike for the UMi model which does not cater for as large variations between BS and MS antenna heights. 
\subsubsection{5GCM UMi LOS Probability Model \cite{GLOBECOMWP1}}
The 5GCM provides \emph{two} LOS probability models. The first one is identical in form to the 3GPP TR 38.901 outdoor model, yet with slightly different curve--fit
parameters ($d_1$ and $d_2$). The second LOS probability model is known as the \emph{New York University (NYU) Squared Model} \cite{RAPPAPORT1}, which improves the accuracy of the $d_1/d_2$ model by including a \emph{square} on the whole term. Though not standardized, the NYU model was developed using a much finer resolution intersection test than the one used by 3GPP TR 36.873/38.901, and uses a real--world database from downtown New York City \cite{RAPPAPORT1}. For UMi scenarios, the 5GCM $d_1/d_2$ model has a slightly smaller mean square error (MSE), but the NYU squared model has a more rapid decay over distance for urban clutter \cite{RAPPAPORT1}. The exact expressions for the NYU model is as follows \cite{GLOBECOMWP1,GLOBECOMWP2}:
\vspace{-1pt}
\begin{align}
   \nonumber
   P\hspace{1pt}_{\textrm{LOS}}^{\textrm{NYU,UMi}}
   \left(d\hspace{1pt}\right)&=\min
    \left(\frac{d_1}{d},1\right)
    \left(1-e^{-d/d_2}\right)\\[3pt]\label{5GCM1}
    &+e^{\hspace{1pt}
    \left(-d/d_2\right)^2}.\\[-22pt]
    &\nonumber
\end{align}
Note that the constants $d_{1}$ and $d_{2}$ are identical to those quoted for the 3GPP models and the 5GCM UMa LOS probability model is identical to the 3GPP TR 36.873/38.901 channel model. For further information, readers are referred to \cite{3GPP25996,3GPP2,3GPP3,GLOBECOMWP1,GLOBECOMWP2} and references therein. 
\vspace{-9pt}
\subsection{Oxygen and Molecular Absorption}
\begin{figure}[!t]
    \centering
    \hspace{-10pt}
    \includegraphics[width=8.7cm]{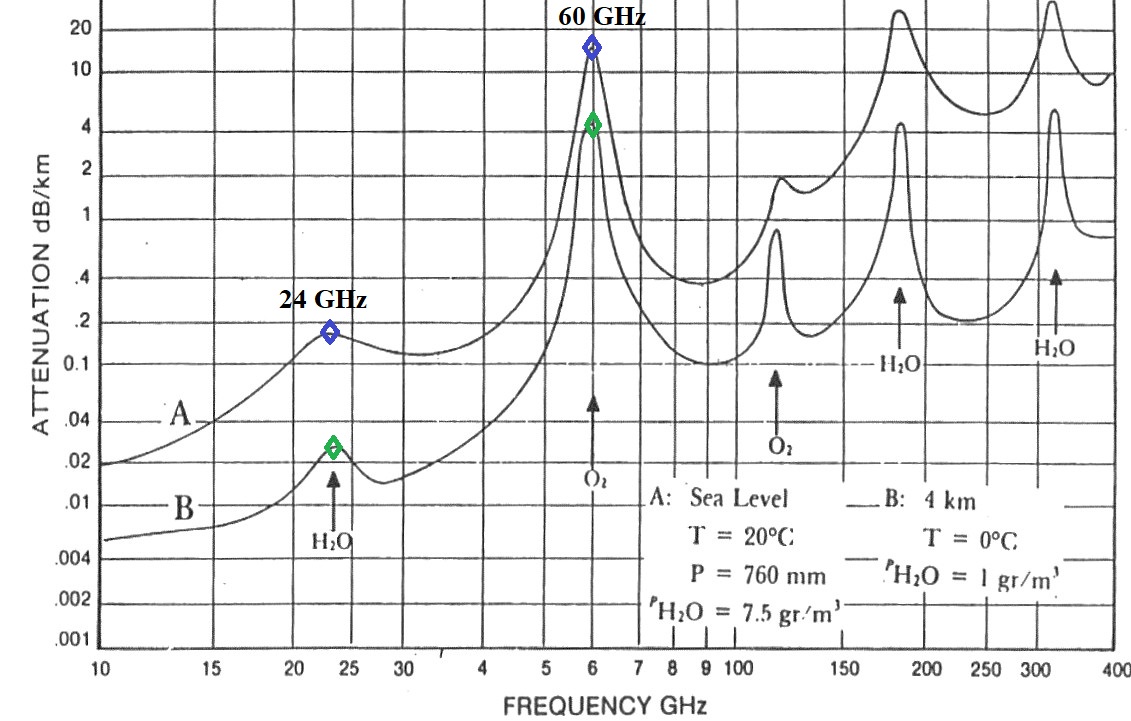}
    \caption{Average atmospheric absorption vs. carrier frequency from 10 GHz to 400 GHz. The two curves denote the \emph{sea--level} attenuation and \emph{attenuation at 4 km}, where various peaks and troughs are observed for oxygen and water sensitive regions. The figure is cited and modified from its source in \cite{FCCreport}.}
    \label{FCCFig}
    \vspace{-8pt}
\end{figure}
Since majority of the standardized cellular systems operate in bands below 6 GHz, oxygen and water vapour absorption has not been under consideration in the design of standardized propagation models. However, for IMT--2020/3GPP 5G--NR systems, the amalgamation of mmWave bands together with bands below 6 GHz makes its consideration important for larger distances. The transmitted wavefronts encounter additional losses due to the absorption of oxygen molecules, water vapour and other gaseous constituents present in the air. These losses are much more pronounced at certain frequencies than others as they coincide with the mechanical resonant frequencies of the gas molecules \cite{FCCreport}. Figure~\ref{FCCFig} shows several peaks that occur due to absorption of the radio signals by water vapour and oxygen molecules. At the resonant frequencies, absorption results in much higher  attenuation, and as a result impacts the net received power for a given link distance. As marked with blue and green diamonds on Fig.~\ref{FCCFig}, the peaks at 24 GHz and 60 GHz are relevant for the case of IMT--2020/5G--NR systems, since the 24 GHz band is used for earth exploration satellite system (EESS) passive sensors that are used to predict water vapour content in the atmosphere, and in turn used for global weather prediction. The 60 GHz band is used for short range wireless local area networks where the link range is not an issue. The spectral regions in between the absorption peaks provide the so--called \emph{low--loss windows}, where propagation can more
readily occur. These transmission windows are at around 24.25-28 GHz, 37--43.5 GHz, 45.5--47 GHz, 47.2--48.2 GHz and 66--71 GHz, respectively, and are identified by WRC 2019 as new frequency bands for IMT2020/5G--NR systems as well as looking beyond 5G systems \cite{WRC2019}. 3GPP TR 38.901 presents a standardized model for oxygen absorption loss which is applied to the cluster responses. Assuming a carrier frequency $f$, the additional loss in dB for cluster $n$ is modelled as 
\vspace{-2pt}
\begin{equation}
    \label{oxygenabsorptionloss}
    L_{n}^{\textrm{oxygen}}\left(f\right)=\frac{\alpha\left(f\right)}{1000}\left[\hspace{1pt}d+c\left(\tau_n+\tau_\Delta\right)\hspace{1pt}\right], 
\end{equation}
where $\alpha(f)$ is the frequency dependent oxygen absorption loss in dB/Km for frequency $f$ as shown in Table 7.6.1--1 in \cite{3GPP3}. Furthermore, $d$ is the 3D link distance, $c$ is the speed of light, $\tau_n$ is the mean cluster delay for the $n$--th cluster and $\tau_\Delta$ denotes the minimum delay of all MPCs belonging to cluster $n$. For larger bandwidths, the parameter of $\alpha(f)$ is replaced by $\alpha(f+\Delta{}f)$, with its value obtained from Sec.~7.6.1 of \cite{3GPP3}. 
\vspace{-12pt}
\subsection{Vegetation Attenuation}
\vspace{-1pt}
Besides the attenuation due to atmospheric effects, the radio signal may also experiences other kinds of attenuation, such as the attenuation as a result of surrounding vegetation. Despite the many pre--standardization measurement--based results on attenuation due to vegetation (see e.g., \cite{SHAFI1} for references), the wide range of foliage types has made it difficult to develop a generalized prediction procedure which can be standardized across wide frequency bands and bandwidths \cite{ITURvege}. Attempts during the past six years have been made to integrate the various published results into standardization. For instance, the work of \cite{Erceg} was included in IEEE 802.16. Signal attenuation with/without vegetation, models for delay spread, Doppler spread, and polarization changes are the factors discussed in \cite{Erceg}. Another key contribution came from \cite{Eric Pelet} in which the authors observe the impact of foliage at the 2.5 GHz band, which was later discussed in standardization but was not included in standardized models. The main findings of \cite{Eric Pelet} is summarized as follows: 1) On a calm day with less than 5 km/h wind, the presence of tree foliage did not cause the strength of the LOS path to change significantly but in the presence of  winds ranging from 10 km/h, to 25 km/h, around 22 dB fades can be observed. 2) The signal power loss through the foliage is also measured to be about the same with and without rain. However, under intense rainfall and no wind, up to 13 dB fades were observed but this increased to 33 dB after rain. Models for tree trunk and leaf attenuation are also given in \cite{ITURvege}. More recently, for 5G--NR systems, the vegetation attenuation at mmWave frequencies is also substantial and is shown to increase with the length of the traveling path through the foliage, though this attenuation saturates for longer distances as paths around the canopies become more dominant \cite{SHAFI1,Schwering}. The resulting attenuation coefficients depend on the vegetation type, season of the year, as well as BS and MS elevation. Keeping in mind the above, we state that there are no specific models for attenuation due to vegetation in the 5G--NR standardized models by the ITU--R/3GPP \cite{ITURM2135,ITU3}. 

\vspace{-8pt}
\subsection{Outdoor--to--Indoor Penetration Loss}
\vspace{-1pt}
\label{OutdoortoIndoorPenetrationLoss}
Standardized models for 2G, 3G and 4G systems typically factor the outdoor--to--indoor 
penetration loss within the pathloss calculation from an outdoor BS to indoor MSs \cite{ITUR,3GPP25996,ITURM2135,3GPP3}. The general trend of such a penetration loss model is typically given by 
\vspace{-5pt}
\begin{equation}
    \label{O2IPenetrationLoss}
    \textrm{PL}=\textrm{PL}_{\textrm{b}}+\textrm{PL}_{\textrm{tw}}+\textrm{PL}_{\textrm{in}}, 
    \vspace{-4pt}
\end{equation}
where $\textrm{PL}_{\textrm{b}}$ is the basic outdoor pathloss model obtained from either 
\cite{ITUR,3GPP25996,ITURM2135,3GPP3}, where $d_{\textrm{3D}}$ is replaced by 
$d_{\textrm{3D-out}}+d_{\textrm{3D-in}}$, $\textrm{PL}_{\textrm{tw}}$ is the building penetration loss through the external wall and $\textrm{PL}_{\textrm{in}}$ is the so--called inside loss which is in turn dependent on the depth of the radio signal into the building. Typical values of these parameters is as given in \cite{ITUR,3GPP25996,ITURM2135,3GPP3}. In contrast to the standardized models from 2G to 4G systems, 5G models \cite{ITU2,3GPPTR38900,3GPP2} typically include an extra factor, $\mathcal{N}(0,\sigma_{P}^{2})$, which models the \emph{standard deviation} of the penetration loss. Note that $\textrm{PL}_{\textrm{tw}}$ is characterized in standardized 5G models as 
\vspace{-4pt}
\begin{equation}
    \label{PLtw5G}
    \textrm{PL}_{\textrm{tw}}=\textrm{PL}_{\textrm{npi}}-10\log
    \left(\sum\limits_{i=1}^{N}\hspace{2pt}p_{i}\hspace{1pt}
    10^{\frac{L_{\hspace{1pt}\textrm{material,i}}}{-10}}\right), 
    \vspace{-3pt}
\end{equation}
where $\textrm{PL}_{\textrm{npi}}$ is an additional loss added to the external wall loss to account for the non--perpendicular incidence. Moreover, the general form of  $L_{\textrm{material,i}}=a_{\textrm{material,i}}+b_{\textrm{material,i}}f$ is the penetration loss of material $i$, where the example values are found in Table 7.4.3--1 of \cite{3GPP2}, $p_{i}$ is the proportion of the $i$--th material, where $\sum_{i=1}^{P}p_{i}=1$ with a total number of $N$ materials. 5G-NR systems are calibrated against two model types, categorized as \emph{low--loss} and \emph{high--loss} models both having differences in the pathloss computations through the wall, net indoor pathloss and standard deviation of the loss, as depicted in \cite{3GPP2}. For the low--loss model, a 70\%:30\% concrete:standard glass material composition is considered, while for the high--loss model, a 70\%:30\% infrared reflective glass:concrete is considered based on the formulations presented in \cite{3GPP2}. Both the low--loss and high--loss models are applicable to UMa and UMi street canyon situation. \emph{The aggregate impact of outdoor--to--indoor penetration loss is added to the shadow fading realization in the logarithm domain.}
\begin{figure}[!t]
    \vspace{-10pt}
    \includegraphics[width=8.8cm]{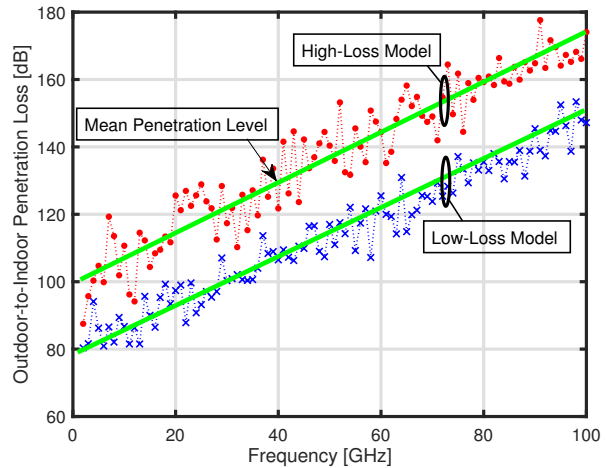}
    \vspace{-15pt}
   \caption{3GPP TR 38.901 quoted outdoor--to--indoor penetration loss model as a function of the carrier frequency for an UMi scenario with a 3D link distance of 25 m. Both the low--loss and high--loss models are depicted with the mean penetration levels with varying frequency.}
   \label{OutdoortoIndoorPenetrationLoss}   
   \vspace{-10pt}
\end{figure}
Figure~\ref{OutdoortoIndoorPenetrationLoss} depicts the net outdoor-to-indoor penetration loss as a function of the carrier frequency from 2 to 100 GHz. Two important observations can be made from the presented result. Firstly, irrespective of the model type (i.e., low--loss or high--loss), at a link distance of 25 m, the penetration loss linearly increases with increasing frequency. This trend can be seen from the instantaneous penetration loss curves in red and blue colors, as well as from the mean penetration loss levels shown by the green lines for both loss model types. Secondly, for the high--loss model, due to the large standard deviation of the excess loss mimicking the interacting incident angles for the chosen material composition, larger variations in the net penetration loss are observed. Such an effect is modelled via the $\mathcal{N}$ term in \eqref{O2IPenetrationLoss} with mean zero and standard deviation of 6.5 dB, in contrast to 4.4 dB for the low--loss model. 

\vspace{-13pt}
\subsection{Blockage Modelling}
\label{BlockageModelling}
\vspace{-1pt}
Like the other additional modelling features, earlier generation cellular systems did not explicitly standardize blockage modelling, apart from the conventional lognormal (a Gaussian random variable in the dB domain) shadowing which factors in \emph{environmental shadowing}, which may be experienced as the MS moves along a trajectory. In such a case, large power variations which occur as the MS moves into and out of regions that are covered by the BS via different propagation mechanisms are modelled. We note that environmental shadowing is typically shown to be spatially correlated across 50--100 m in bands below 6 GHz (see e.g., \cite{3GPP3,ITURM2135}). The well cited work of \cite{Gudmundson} is the sole model which is standardized for modeling correlation shadowing. Unlike sub--6 GHz frequencies, the lack of diffraction efficiency at mmWave frequencies makes shadowing significantly more pronounced. In addition to environmental shadowing, recently standardized models for 5G--NR systems also explicitly model shadowing \emph{induced by environmental objects}, i.e., when a MPC is \emph{blocked} by an object or a human being, as well as \emph{self--shadowing} induced by the person holding the MS. The later naturally depends on rotation and change of hold (e.g., MS to ear vs. in front of torso mode) of the MS. The 3GPP/ITU--R have standardized two blockage models in TR 38.901/M.2412, known as Blockage Model A and Blockage Model B \cite{3GPP2,ITU2}. Both approaches are designed to serve their own use cases. Model A is applicable for generic human and vehicular blockages, where an iterative step-wise procedure for modifying the small--scale fading cluster parameters is presented. Parameters describing the blockage region are defined for both indoor and outdoor scenarios, such as UMi, UMa, RMa and InH, respectively. Model B applies a more specific geometrical model to the methodology outlined in Model A. In particular, the total number of blockers, their vertical and horizontal extensions, and their relationship as a function of distance from the blockage point to the MS are defined in Table 7.6.4.2--5 of \cite{3GPP2}. Furthermore, two different geometry--based methods to determine the blockage attenuation per--path are given, namely for LOS and NLOS paths, respectively. We encourage the reader to further refer to \cite{3GPP2,ITU2} for a more detailed overview of both the proposed blockage models. 
\vspace{-13pt}
\subsection{Spatial Consistency Modelling}
\label{subsec:sptial_consistency}
\begin{figure}[!t]
\vspace{-12pt}
\begin{center}
\includegraphics[width=8.7cm]{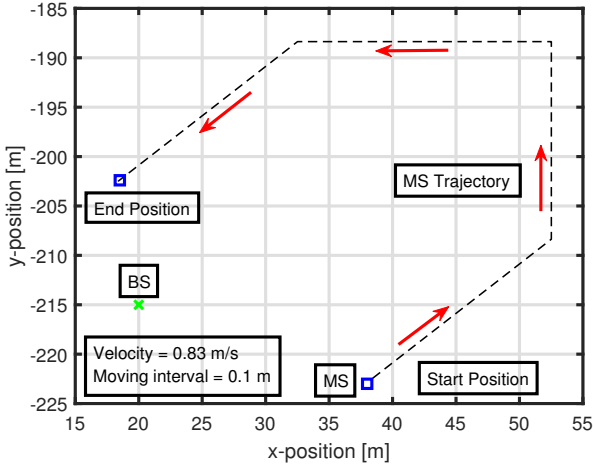}
\vspace{-7pt}
\caption{Investigated scenario where the MS communicates with the BS while moving along a trajectory shown by red arrows with a predefined velocity and moving interval.}
\label{scenarioA}
\end{center}
\vspace{-23pt}
\end{figure}
\begin{figure}[!t]
\vspace{-13pt}
\begin{center}
\includegraphics[width=8.7cm]{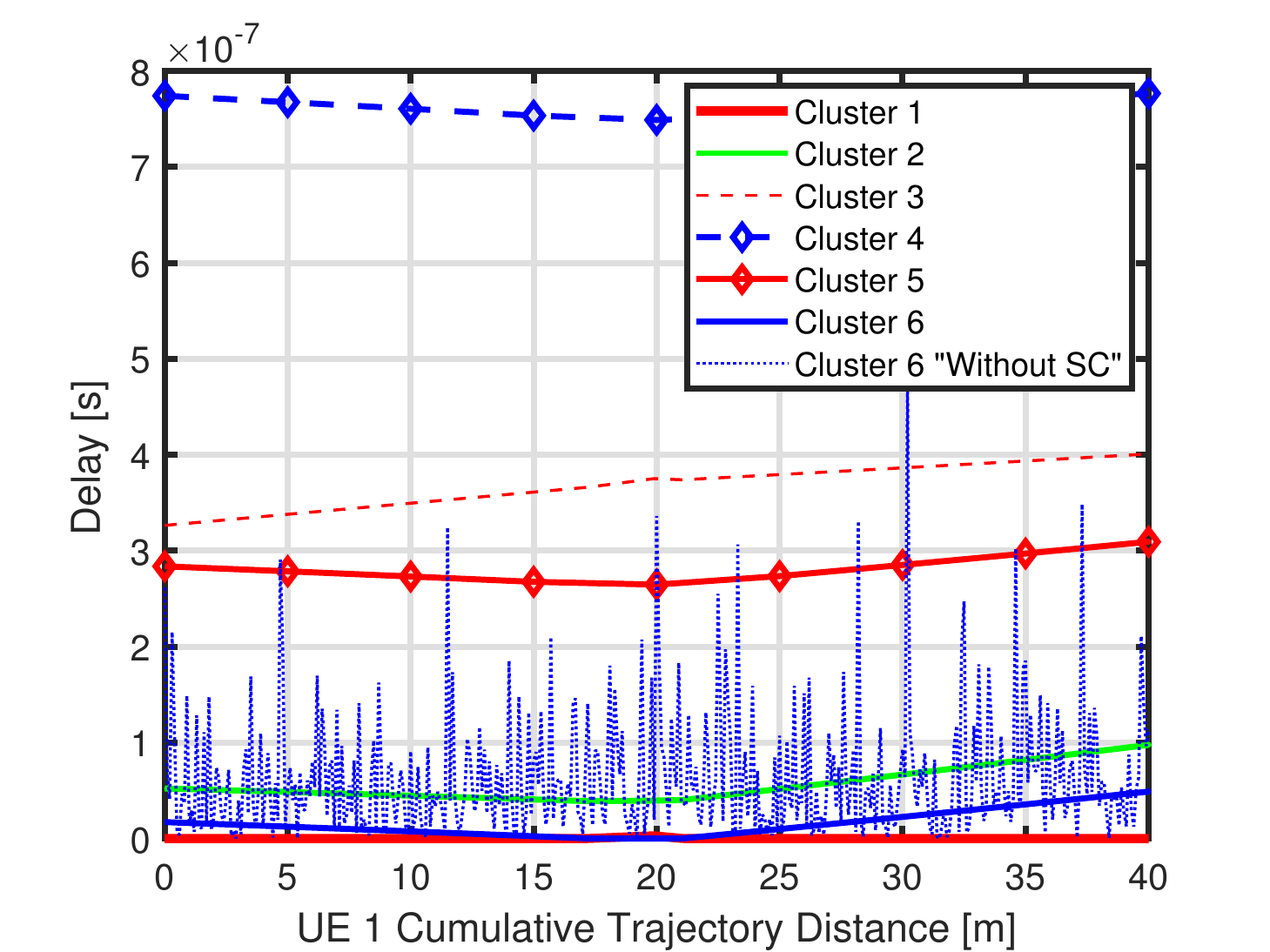}
\vspace{-6pt}
\caption{Cluster delays with and without spatial consistency vs. cumulative distance of MS trajectory.}
\label{delays}
\end{center}
\vspace{-15pt}
\end{figure}
Spatial consistency is regarded as a novel mandatory feature of 5G--NR propagation models. Earlier standardized models by the 3GPP/ITU--R are based on the concept of a \emph{drop}. In the standardization terminology, a drop is an instantaneous channel segment which represents a period of \emph{quasi–-stationarity}. In a given drop, the large–scale and small-scale parameters needed to generate the overall channel impulse response obey their respective distributions. However, between multiple drops, the channel parameters have no continuity, and \emph{independence} is assumed. In reality, it is naturally desirable that propagation parameters maintain continuity across multiple realizations. This is particularly important when the MS moves along a trajectory or when there is movement of scatterers influencing the channel impulse response. Unlike the 3GPP/ITU--R models from 3G and 4G, the COST 259/273/2100 channel models have a long history of implementing the concept of spatial consistency via the concept of random visibility regions of clusters. Spatial consistency modelling is particularly important in 5G--NR systems for predicting the system level performance with the newly proposed beam tracking algorithms of 3GPP Release 15 and 16, respectively \cite{SHAFI2,TatariaEuCAP}. The ITU--R M.2412 and 3GPP TR 38.901 models both define two procedures for spatially consistent mobility modelling, namely Spatial Consistency Model I/Model A, which we denote as SC–-I, and Spatial Consistency Model II/Model B, which we denote as SC--II. The SC--I model applies an iterative algorithm to update the propagation parameters with a restricted moving distance of the MS between consecutive channel realizations which is limited by the correlation distance of the parameters. In SC–-II, according to the location of the MS, spatially consistent channel parameters are obtained separately, where the cluster delays and angles are generated with a modified procedure. As an example, we provide one realization of SC–-I. Firstly, the time axis information described in \cite{3GPP2,ITU2} is added into the simulation procedure. The initial MPC amplitude, delay, and angular parameters are generated according to the same procedure as without SC. Assuming a MS moves along a trajectory at a speed, $v$, the moving distance will be limited within one meter in a short time epoch, $\Delta{}t$. Then, for each $\Delta{}t$ interval, the delays, powers and angles will be updated with the method in \cite{ITU2}. Finally, these updated parameters will be used to generate the overall channel impulse response. 
\begin{figure}[!t]
\vspace{-5pt}
\begin{center}
\vspace{-5pt}
\includegraphics[width=8.7cm]{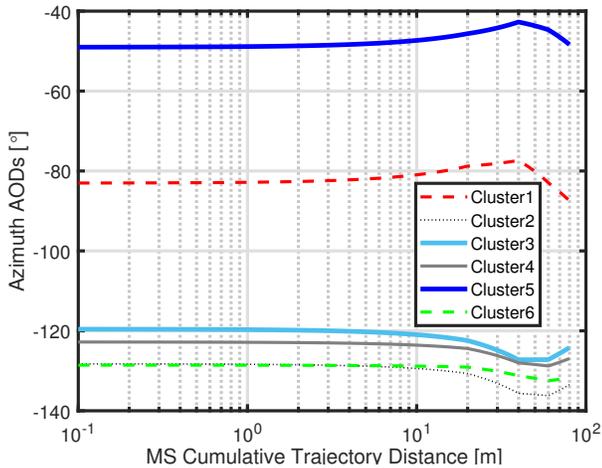}
\vspace{-6pt}
\caption{Spatially consistent azimuth AODs vs. cumulative distance of MS trajectory.}
\label{aods}
\end{center}
\vspace{-5pt}
\end{figure}
\begin{figure}[!t]
\vspace{-6pt}
\begin{center}
\includegraphics[width=8.7cm]{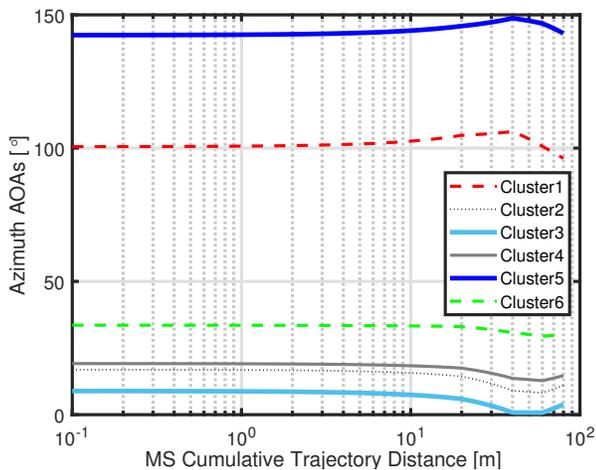}
\vspace{-5pt}
\caption{Spatially consistent azimuth AOAs vs. cumulative distance of MS trajectory.}
\label{aoas}
\end{center}
\vspace{-20pt}
\end{figure}

We consider the scenario presented in Fig.~\ref{scenarioA}. The BS (marked with a green cross) is located approximately 20 m away from the start position of the MS (marked with a blue square). The MS is moving along the trajectory shown by black dashed line and red arrows from its start position to finish position as depicted in Fig.~\ref{scenarioA}. The MS is set to move at the interval of 0.1 m with a velocity of 0.83 m/s (approximately 3 km/h). The propagation channel is modelled as in the outdoor UMi model of 3GPP TR 38.901. The specific cluster characteristics in both azimuth and elevation domains follow the 3GPP defined statistical distributions. The update distance of MS is set to 0.1 m, so that we are able to accurately capture variations with sufficiently high resolution in the propagation channel in comparison to the 15 m correlation distance of large--scale parameters defined by the 3GPP. In order to maintain clarity and to minimize cluttering the results, our focus is confined on a subset of propagation parameters, since the conclusions drawn are also valid for other parameters. The spatially consistent delays belonging to the \emph{first six clusters} for the UMi environment are depicted in Fig.~\ref{delays}. From the figure, it can be readily observed that spatial consistency makes the cluster delays evolve continuously and smoothly through the entire MS track. \emph{In order to analyze the difference between spatially consistent vs. spatially inconsistent delays, we take the example of cluster 2, where the delay response without spatial consistency is shown. Here, one can notice a clear difference as the delays tend to fluctuate abruptly and are discontinuous from one channel segment to the next, illustrating the classical drop--based concept.} Similar effects can be observed on the azimuth AODs and AOAs, which are demonstrated in Figs.~\ref{aods}  and \ref{aoas}, respectively. Here one can also observe the continuous evolution of the cluster angles relative to trajectory distance induced by the SC--I procedure. For further discussions, the readers are asked to see \cite{SHAFI2,TatariaEuCAP}. 

\vspace{-12pt}
\subsection{Correlation Modelling for Multi--frequency Simulations}
Standardized models of 3G and 4G systems lacked sufficient measurement data and statistical validation to conclude if any correlation was being induced across the different large--scale and small--scale parameters in given environment, at a given frequency band. However, standardized models for 5G--NR systems have devised a methodology to generate and analyze the correlation in propagation parameters across a set of frequencies. Specifically, section 7.6.5 of \cite{3GPP2} reports a step--wise process to correlate the amplitude, delays, angles, LOS probability, per--cluster shadowing, and antenna array patterns as a function of the center frequency and bandwidth. 

\vspace{-10pt}
\section{Conclusions}
\vspace{-1pt}
Summarizing over 40 years of history of channel models and their standardization is a formidable task, and that too within the limited space of a journal paper. In this paper, we have attempted such a description, together with the key technical features of the models. Due to the wide scope of the contents, we have assumed the reader is familiar with the fundamentals of propagation. A very significant evolution of all aspects of propagation modeling has happened over the last 40 years. The complexity of models has also significantly evolved from simple pathloss to double--directional impulse response for antenna arrays used for massive MIMO and beamforming. The use of mmWave bands requires special consideration of special features like indoor--to--outdoor penetration loss, oxygen absorption, vegetation loss, and blockages. All these aspects and more are discussed in this paper. Finally, the standardized models given in \cite{3GPP2} and \cite{ITU2} allow the reader to simulate wireless channels over the frequency range of 900 MHz to 100 GHz.

\vspace{-19pt}
\section*{Author Biographies}
\normalsize

\textbf{Harsh Tataria} received his Ph.D. degree from Victoria University of Wellington, New Zealand, in 2017. From March 2017 to October 2018, he was a Postdoctoral Fellow at Queen's University Belfast, UK. Since November 2018, he is with Lund University, Sweden. His research interests include real--time channel measurements and models. 

\textbf{Katsuyuki Haneda} received the D.Eng degree from the Tokyo Institute of Technology, Tokyo, Japan, in 2007. He is presently an Associate Professor with the School of Electrical Engineering, Aalto University, Espoo, Finland. His current research activity includes radio frequency instrumentation, measurements and modeling, millimeter-wave radios, wireless for medical and post--disaster scenarios and in--band full-duplex radio technology. Dr. Haneda has been an Associate Editor for the IEEE Transactions on Antennas and Propagation from 2012--2016, and an Editor for the IEEE Transactions on Wireless Communications from 2013--2018.

\textbf{Andreas F. Molisch} received the Dipl.Ing., Ph.D., and Habilitation degrees from the Technical University of Vienna, Vienna, Austria, in 1990, 1994, and 1999, respectively. He spent the next 10 years in industry, at FTW, AT\&T (Bell) Laboratories, and Mitsubishi Electric Research Labs (where he rose to Chief Wireless Standards Architect). In 2009 he joined the University of Southern California (USC) in Los Angeles, CA, where he is now the Solomon Golomb--Andrew and Erna Viterbi Chair. His current research interests are the measurement and modeling of mobile radio channels, multiantenna systems, wireless video distribution, ultrawideband communications and localization, and novel modulation formats. 
He has authored, co--authored, or edited four books among them the textbook Wireless Communications, 20 book chapters, more than 250 journal articles, more than 350 conference papers, as well as more than 80 patents and 70 standards contributions. He is a Fellow of the National Academy of Inventors, AAAS, IEEE, and IET, an IEEE Distinguished Lecturer, a Member of the Austrian Academy of Sciences, and a recipient of many awards. 

\textbf{Mansoor Shafi} received the B.S. degree from the University of Engineering and Technology, Lahore, in 1970, and the Ph.D. degree from the University of Auckland, Auckland, New Zealand, in 1979, both in Electrical Engineering. He is currently with Spark NZ and is a Telecom Fellow and an Adjunct Professor with the School of Engineering, Victoria University. He has authored and coauthored more than 150 journal papers in the area of wireless communications. He has shared Best Tutorial Paper Award from the IEEE Communications Society in 2004 and the IEEE Donald G. Fink Award in 2011. He was awarded member of the New Zealand Order of Merit in the Queen's Birthday Honors 2013 for services to wireless communications. He is a Life Fellow of the IEEE.  

\textbf{Fredrik Tufvesson} received the Ph.D. degree from Lund University, Lund, Sweden, in 2000. After two years at a startup company, he joined the Department of Electrical and Information Technology, Lund University, where he is currently Professor of radio systems. He has authored around 90 journal papers and 140 conference papers. His main research interests
include the interplay between the radio channel and the rest of the communication system with various applications in 5G systems, such as massive MIMO, mmWave communication, vehicular communication, and radio--based positioning. He recently received the Neal Shepherd Memorial Award for the best propagation paper in the IEEE Transactions on Vehicular Technology and the IEEE Communications Society Best Tutorial Paper Award. He is a fellow of the IEEE. 
\end{document}